\documentclass[11pt]{article}
\usepackage{natbib, graphicx, setspace, amsmath, amssymb, subfigure, url, multirow, booktabs, verbatim, bm,threeparttable,algorithm, color}
\usepackage{amsthm}

\allowdisplaybreaks[3]
\begin{document}

\setlength{\textheight}{575pt}
\setlength{\baselineskip}{23pt}

\title{Structured Gene-Environment Interaction Analysis}

\author{Mengyun Wu$^{1,3}$, Qingzhao Zhang$^2$, Shuangge Ma$^3$ \\ \\
$^1$School of Statistics and Management, Shanghai University of Finance and Economics\\
$^2$School of Economics and Wang Yanan Institute for Studies in Economics, Xiamen University\\
$^3$Department of Biostatistics, Yale University\\
 \\
email: shuangge.ma@yale.edu}

\date{}
\maketitle

\begin{abstract}
For the etiology, progression, and treatment of complex diseases, gene-environment (G-E) interactions have important implications beyond the main G and E effects. G-E interaction analysis can be more challenging with the higher dimensionality and need for accommodating the ``main effects, interactions'' hierarchy. In the recent literature, an array of novel methods, many of which are based on the penalization technique, have been developed. In most of these studies, however, the structures of G measurements, for example the adjacency structure of SNPs (attributable to their physical adjacency on the chromosomes) and network structure of gene expressions (attributable to their coordinated biological functions and correlated measurements), have not been well accommodated. In this study, we develop the structured G-E interaction analysis, where such structures are accommodated using penalization for both the main G effects and interactions. Penalization is also applied for regularized estimation and selection. The proposed structured interaction analysis can be effectively realized. It is shown to have the consistency properties under high dimensional settings. Simulations and the analysis of GENEVA diabetes data with SNP measurements and TCGA melanoma data with gene expression measurements demonstrate its competitive practical performance.
\end{abstract}


\section{Introduction}

Beyond the main genetic (G) and environmental (E) effects, gene-environment (G-E) interactions have been shown to be fundamentally important for the etiology, progression, prognosis, and response to treatment of many complex diseases. In the past decade, a long array of statistical methods have been developed for G-E interaction analysis and can be roughly classified as marginal analysis (under which one G measurement is analyzed at a time) and joint analysis (under which a large number of G measurements are analyzed in a single model). For relevant discussions, we refer to \cite{Thomas10,Wu15,Wu18,Shim18} and other published studies. Compared to marginal analysis, joint analysis may better describe disease biology (that is, phenotypes and outcomes of complex diseases are associated with the combined effects of multiple genetic factors) and have attracted extensive attention in recent literature.

Joint G-E interaction analysis is challenging with the high data dimensionality. For estimation and also to screen out noises and identify important G-E interactions and main G effects, regularized estimation has been routinely conducted. Among the available techniques, penalization has been adopted in many of the recent studies. See \cite{Wu15, Wu18} and references therein. Another challenge comes from the need to respect the ``main effects, interactions'' hierarchy \citep{Bien13,Hao17,Hao18,She18}. Under the context of G-E interaction analysis with low-dimensional E variables, this hierarchy postulates that an interaction term cannot be identified, if the corresponding main G effect is not identified. With this hierarchy, ``straightforward'' penalizations are insufficient. Several penalization techniques have been developed in recent literature to respect this hierarchy \citep{Liu13,Wu18dissecting}.

A common limitation shared by many of the existing G-E interaction studies is that the structures of G measurements have not been well accounted for. Consider for example single nucleotide polymorphism (SNP) data. When SNPs are densely measured, those physically close are often in high linkage disequilibrium (LD) and likely o have similar biological functions or statistical effects \citep{Reich01}. Here, there is an adjacency structure which arises from the physical adjacency of SNPs on chromosomes. As another example, consider gene expressions. Recent studies have shown that with coordinated biological functions and correlated measurements, gene expressions can be effectively described using a network structure \citep{Barabasi11}. More details are provided below. Note that for other types of omics measurements, there are also underlying structures, although the construction of such structures may vary across data types.

In the high-dimensional analysis of main G effects, a few structured analysis approaches have been developed to accommodate the underlying structures in estimation and selection. Consider the adjacency structure of SNPs (and other densely measured G factors). Available penalization approaches include the fused lasso \citep{Tibshirani05}, smooth lasso \citep{Hebiri11}, smoothed group lasso \citep{Liu12}, spline lasso \citep{Guo16}, and others. When gene expressions (and other G measurements) are described using network structures, network-constrained regularized estimation has been proposed \citep{Michailidis12}. A popular approach is the network Laplacian-based penalization \citep{Li08}. Other network-structured penalization methods include the adaptive network-constrained regularization \citep{Li10}, TLP-based penalty for groups of indicators \citep{Kim13}, sparse regression incorporating graphical structures among predictors (SRIG) \citep{Yu16}, and others. Extensive investigations have shown that structured analysis can lead to more accurate and more interpretable identification and estimation of important effects. It is noted that, with similar spirits, structured analysis can also be conducted based on techniques other than penalization. As penalization is adopted in this study, the above literature review has been focused on this specific technique.

In this study, our goal is to conduct structured G-E interaction analysis, under which the structures of G measurements can be effectively accounted for. This has been well motivated by the success of structured analysis in the study of main G effects and a lack of such analysis in G-E interaction analysis. It is noted that this study is much more than a simple extension of the main-G-effect structured analysis. Specifically, in G-E interaction analysis, one G factor manifests multiple effects: its main effect as well as multiple E-interactions. The underlying structures need to be accounted for in the analysis of all these effects. This is further complicated by the ``main effects, interactions'' hierarchy. As a result, significant computational and statistical developments are needed. Also advancing from some of the existing studies, in this study, we accommodate multiple types of underlying structures under one framework.
This unity significantly benefits methodological and statistical developments.
Another advancement is that statistical properties are carefully established, which can provide a more solid ground than in some of the existing studies. Overall, this study can provide an alternative and more effective way for conducting G-E interaction analysis.

\section{Methods}

Consider a dataset with $n$ iid subjects. For the $i$th subject, let $Y_i$ be the response of interest, and $\bm{Z}_{i\cdot} = (Z_{i1}, \cdots, Z_{iq})$ and $\bm{X}_{i\cdot}= (X_{i1}, \cdots, X_{ip})$ be the $q$- and $p$-dimensional vectors of E and G measurements.  First, consider the scenario with a continuous outcome and a linear regression model with the joint effects of all E and G effects and their interactions:
\begin{equation}\label{general_model}
Y_i=\sum_{k=1}^q Z_{ik}\alpha_k+\sum_{j=1}^p X_{ij} \beta_j+\sum_{k=1}^q\sum_{j=1}^p Z_{ik} X_{ij}\eta_{kj} +\varepsilon_i,
\end{equation}
where $\alpha_k$'s, $\beta_j$'s, and $\eta_{kj}$'s are the regression coefficients for the main E, main G, and their interactions, respectively, and $\varepsilon_i$'s are the random errors. We omit the intercept term to simplify notation.

To conveniently respect the ``main effects, interactions'' hierarchical constraint, we conduct the decomposition of $\eta_{kj}$ as $\eta_{kj}=\beta_j \gamma_{kj}$. Then model (\ref{general_model}) can be rewritten as
\begin{eqnarray*}\label{Hier_model}
\nonumber Y_i&=&\sum_{k=1}^q Z_{ik}\alpha_k+\sum_{j=1}^p X_{ij} \beta_j+\sum_{k=1}^q\sum_{j=1}^p Z_{ik} X_{ij} \beta_j \gamma_{kj} +\varepsilon_i\\
&=& \bm{Z}_{i\cdot}\bm{\alpha}+\bm{X}_{i\cdot}\bm{\beta}+\sum_{k=1}^q \bm{W}_{i\cdot}^{(k)}(\bm{\beta}\odot \bm{\gamma}_k) +\varepsilon_i,
\end{eqnarray*}
where $\bm{\alpha}=(\alpha_1,\cdots,\alpha_q)'$, $\bm{\beta}=(\beta_1,\cdots,\beta_p)'$, $\bm{\gamma}_k=(\gamma_{k1},\cdots,\gamma_{kp})'$, $\bm{W}_{i\cdot}^{(k)}=(Z_{ik} X_{i1},\cdots,Z_{ik} X_{ip})$, and $\odot$ is the component-wise product.
Denote $\bm{Y}$ as the $n$-length vector composed of $Y_i$'s, and $\bm{Z}$, $\bm{X}$, and $\bm{W}^{(k)}$ as the $n\times q$, $n\times p$ and $n\times p$ design matrices composed of $\bm{X}_{i\cdot}$'s, $\bm{Z}_{i\cdot}$'s and $\bm{W}_{i\cdot}^{(k)}$'s, respectively.

For estimation and selection of important effects, consider the penalized objective function
\begin{eqnarray}\label{objective_fun}
\nonumber Q_n(\bm{\theta})&=&\frac{1}{2n}\left\| \bm{Y}-\bm{Z}\bm{\alpha}-\bm{X}\bm{\beta}-\sum_{k=1}^q \bm{W}^{(k)}(\bm{\beta}\odot \bm{\gamma}_k)\right\|_2^2+\sum_{j=1}^p \rho(|\beta_j|;\lambda_1,r)+\sum_{j=1}^p \sum_{k=1}^q \rho(|\gamma_{kj}|;\lambda_1,r)\\
&+&\frac{1}{2}\lambda_2\bm{\beta}'\bm{J}\bm{\beta}+
\frac{1}{2}\lambda_2\sum_{k=1}^q\bm{\gamma}_k'\bm{J}\bm{\gamma}_k,
\end{eqnarray}
where $\bm{\theta}=(\bm{\alpha}',\bm{\beta}',\bm{\gamma}')'=
(\bm{\alpha}',\bm{\beta}',\bm{\gamma}_1',\cdots,\bm{\gamma}_q')'$, $||\bm{\nu}||_2$ is the $L_2$ norm of vector $\bm{\nu}$,
$\rho(|\nu|;\lambda_1,r)=\lambda_1 \int_0^{|\nu|}\left(1-\frac{x}{\lambda_1 r}\right)_+dx$ is the minimax concave penalty (MCP), $\lambda_1\geq 0$ and $\lambda_2\geq 0$ are data-dependent tuning parameters, and $r\geq 0$ is the regularization parameter.  $\bm{J}$ is the $p\times p$ matrix that accommodates the structure of G measurements (more details below). The proposed estimate is defined as the minimizer of (\ref{objective_fun}). The nonzero components of $\bm{\beta}$ and $\bm{\beta}\odot \bm{\gamma}_k~(k=1,\cdots, q)$ correspond to the important main G effects and interactions that are associated with the response.

In the objective function, the first term is the lack-of-fit. Each of the first two penalty functions is the sum of $p$ terms. For each of the G factors, penalties are imposed on its main effect as well as interactions. With the decomposition ($\beta_j \gamma_{jk}$), the proposed penalties guarantee that a G-E interaction is not identified if the corresponding main G effect is not identified. Note that here the setting and hence strategy differ from the pairwise interaction analysis studies such as \cite{Choi10}, \cite{Lim15}, and \cite{Hao18}. Specifically, in most G-E interaction analysis, for example as considered in our data examples, the E factors are manually selected based on extensive prior knowledge and have a low dimensionality. As such, there is no need to conduct selection with E effects, and their coefficients are always nonzero. In the literature, there are other ways of achieving the hierarchy, for example, the sparse group MCP \citep{Liu13}. Our exploration suggests that the proposed approach has significant computational advantages.

\noindent{\bf Accommodating the structures of G measurements}
In (\ref{objective_fun}), the underlying structures of G measurements are accommodated using the last two penalty terms. Here for interactions, instead of $\bm{\beta}\odot \bm{\gamma}_k$, we consider the structures of $\bm{\gamma}_k$ which can significantly facilitate theoretical and numerical analysis. Our numerical investigation suggests that two approaches lead to similar results (details omitted). First, consider the following two specific examples.

Consider SNP data. Assume that densely measured SNPs have been sorted according to their physical locations on the chromosomes. Consider the following spline type penalty:
\begin{equation}\label{spline}
\sum_{j=2}^{p-1}\left[(\beta_{j+1}-\beta_{j})-(\beta_{j}-
\beta_{j-1})\right]^2\textrm{ and } \sum_{j=2}^{p-1}\left[(\gamma_{k(j+1)}-\gamma_{kj})-
(\gamma_{kj}-\gamma_{k(j-1)})\right]^2.
\end{equation}
With this penalty, we have $\bm{J}=\bm{H}_{(p-2)\times p}'\bm{H}_{(p-2)\times p}$ with $H_{jj}=H_{j(j+2)}=1, H_{j(j+1)}=-2$, and 0 otherwise. Here $\bm{J}$ is a very sparse matrix. For SNPs as well as their interactions with a specific E factor, this penalty promotes smoothness in a similar way as penalizing second order derivatives in spline-based nonparametric estimation. As a result, adjacent SNPs are promoted to have similar main effects (E-interactions) associated with the response. With main G effects, some alternatives, such as the fused lasso and smooth lasso, promote first-order smoothness, while this penalty promotes second-order smoothness. \cite{Guo16} shows that the spline type penalty can outperform these alternatives. Another advantage of the spline type penalty is that the quadratic form is computationally more manageable than, for example, the absolute-value-based.
It is noted that this study is the first to consider the spline type penalization in the context of G-E interaction analysis.

Consider gene expression data. Following published studies \citep{Shi15}, we first construct the adjacency matrix $\bm{A}=(a_{jl})_{p\times p}$, where $a_{jl}=r^{Pcorr}_{jl}I(|r^{Pcorr}_{jl}|>c^{Pcorr})$ with $r^{Pcorr}_{jl}$ being the Pearson's correlation coefficient between gene expressions $j$ and $l$ and $c^{Pcorr}$ being the cutoff calculated from the Fisher transformation. Note that there are multiple alternatives for constructing the adjacency matrix \citep{Huang11}.
Let $\bm{D}=\textrm{diag}\left(\sum_{l=1}^p |a_{1l}|,\cdots, \sum_{l=1}^p |a_{pl}|\right)$.  We consider
\begin{equation}\label{Laplacian}
\bm{J}=\bm{I}-\bm{D}^{-1/2}\bm{A}\bm{D}^{-1/2},
\end{equation}
where $\bm{I}$ is the $p\times p$ identity matrix. With the cutoff $c^{Pcorr}$, $\bm{J}$ is usually a sparse matrix. This penalty encourages the effects of correlated gene expressions to be similar, which is adjusted by the degree of adjacency. Several recent studies have established the effectiveness of this Laplacian penalization strategy for the analysis of main G effects. However, its adoption in the context of G-E interaction analysis is still lacking.

As can be seen from the above two examples, the definition of $\bm{J}$ needs to be adapted to the specific data settings and may vary across data types. On the other hand, the above definitions can be extended and applied to quite a few other dense and ``non-dense'' cases, making the proposed analysis broadly applicable.

\noindent{\bf Accommodating other response variables}
In the above definition as well as some downstream developments, we use the continuous outcome and linear model as an example. The proposed approach can be extended to other data types/models. For example, in our numerical study, we consider the censored survival outcome and accelerated failure time (AFT) model. Details on this setting are provided in Appendix.

\subsection{Computation}

With fixed tuning parameters, optimization of (\ref{objective_fun}) can be conducted using an iterative coordinate descent (CD) algorithm, which optimizes the objective function with respect to one of the three vectors, $\bm{\alpha}$, $\bm{\beta}$, and $\bm{\gamma}$, at a time and iteratively cycles through all parameters until convergence is reached. The proposed algorithm proceeds as follows:

\noindent \textbf{Step 1} Initialize $t=0$, $\bm{\beta}^{(t)}=\bm{0}$, $\bm{\gamma}^{(t)}=\bm{0}$, and $\bm{\alpha}^{(t)}=(\bm{Z}'\bm{Z})^{-1}\bm{Z}'\bm{Y}$, where $\bm{\alpha}^{(t)}$, $\bm{\beta}^{(t)}$, and $\bm{\gamma}^{(t)}$ denote the estimates of $\bm{\alpha}$, $\bm{\beta}$, and $\bm{\gamma}$ at iteration $t$, respectively.

\noindent \textbf{Step 2} Update $t=t+1$. With $\bm{\gamma}$ and $\bm{\alpha}$ fixed at $\bm{\gamma}^{(t-1)}$ and $\bm{\alpha}^{(t-1)}$, optimize (\ref{objective_fun}) with respect to $\bm{\beta}$. Let $\tilde{\bm{Y}}^{(t)}=\bm{Y}-\bm{Z}\bm{\alpha}^{(t-1)}$ and $\tilde{\bm{X}}^{(t)}=\bm{X}+\sum_{k=1}^q \bm{W}^{(k)}\odot\left(\bm{1}_{n\times 1}\left(\bm{\gamma}_{k}^{(t-1)}\right)'\right)$ with $\bm{1}_{n\times 1}=(1,\cdots,1)_{n \times 1}$. Then
\begin{eqnarray*}\label{beta}
\bm{\beta}^{(t)}=\textrm{argmin} \frac{1}{2n}\left\|\tilde{\bm{Y}}^{(t)}-\tilde{\bm{X}}^{(t)}
\bm{\beta}\right\|_2^2+\sum_{j=1}^p \rho(|\beta_j|;\lambda_1,r)+\frac{1}{2}\lambda_2\bm{\beta}'
\bm{J}\bm{\beta}.
\end{eqnarray*}
For $j=1,\cdots,p$, carry out the following steps sequentially.
\begin{enumerate}
\item []\textbf{Step 2.1} Compute
\begin{equation*}
\bm{res}_{-j}^{(t)}=\tilde{\bm{Y}}^{(t)}-\sum_{l=1}^{j-1}
\tilde{\bm{X}}_{l}^{(t)}\beta_l^{(t)}-\sum_{l=j+1}^p
\tilde{\bm{X}}_{l}^{(t)}\beta_l^{(t-1)},~
\chi_j^{(t)}=\frac{1}{n}\left(\bm{\tilde{X}}_{j}^{(t)}\right)'
\bm{\tilde{X}}_{j}^{(t)},
\end{equation*}
\begin{equation*}
\varphi_j^{(t)}=\frac{1}{n}\left(\bm{\tilde{X}}_{j}^{(t)}\right)'
\bm{res}_{-j}^{(t)},~
\Delta_j^{(t)}=\sum_{l=1}^{j-1} \beta^{(t)}_l J_{jl}+\sum_{l=j+1}^p \beta^{(t-1)}_l J_{jl}.
\end{equation*}
\item []\textbf{Step 2.2} Update the estimate of $\beta_j$ as
\[\beta_j^{(t)}=\left\{\begin{array}{ll}
\frac{\textrm{ST}\left(\varphi_j^{(t)}-\lambda_2\Delta_j^{(t)},
\lambda_1\right)}{\chi_j^{(t)}+\lambda_2J_{jj}-\frac{1}{r}}, & \left|\varphi_j^{(t)}-\lambda_2\Delta_j^{(t)}\right|\leq\lambda_1r
\left(\chi_j^{(t)}+\lambda_2 J_{jj}\right)\\
\frac{\varphi_j^{(t)}-\lambda_2\Delta_j}{\chi_j^{(t)}+\lambda_2J_{jj}}, & \left|\varphi_j^{(t)}-\lambda_2\Delta_j^{(t)}\right|>\lambda_1r
\left(\chi_j^{(t)}+\lambda_2J_{jj}\right)
\end{array}\right.,\]
where $\textrm{ST}(\nu,\lambda_1)=sgn(\nu)(|\nu|-\lambda_1)_+$ is the soft-thresholding operator.
\end{enumerate}

\noindent \textbf{Step 3} With $\bm{\beta}$ and $\bm{\alpha}$ fixed at $\bm{\beta}^{(t)}$ and $\bm{\alpha}^{(t-1)}$, optimize (\ref{objective_fun}) with respect to $\bm{\gamma}$. Let $\breve{\bm{Y}}^{(t)}=\bm{Y}-\bm{Z}\bm{\alpha}^{(t-1)}-\bm{X}
\bm{\beta}^{(t)}$ and $\left(\tilde{\bm{W}}^{(k)}\right)^{(t)}=\bm{W}^{(k)}\odot \left(\bm{1}_{n\times 1}\left(\bm{\beta}^{(t)}\right)'\right)$. Then
\begin{eqnarray*}\label{gamma}
\left(\bm{\gamma}_1^{(t)},\cdots,\bm{\gamma}_q^{(t)}\right)=
\textrm{argmin} \frac{1}{2n}\left\|\breve{\bm{Y}}^{(t)}-\sum_{k=1}^q
\left(\tilde{\bm{W}}^{(k)}\right)^{(t)}\bm{\gamma}_k\right\|_2^2+
\sum_{k=1}^q\sum_{j=1}^p \rho(|\gamma_{kj}|;\lambda_1,r)+\frac{1}{2}\lambda_2\sum_{k=1}^q
\bm{\gamma}_k'\bm{J}\bm{\gamma}_k.
\end{eqnarray*}
For $k=1,\cdots,q$ and $j\in \left\{j:\beta_j^{(t)}\neq0,j=1,\cdots,p\right\}$, conduct estimation similar to Steps 2.1 and 2.2.

\noindent \textbf{Step 4} Compute $\bm{\alpha}^{(t)}=(\bm{Z}'\bm{Z})^{-1}\bm{Z}'\left(\bm{Y}-
\bm{X}\bm{\beta}^{(t)}-\sum_{k=1}^q \bm{W}^{(k)}\left(\bm{\beta}^{(t)}\odot \bm{\gamma}_k^{(t)}\right)\right)$.

\noindent \textbf{Step 5} Repeat Steps 2-4 until convergence. In our numerical study, convergence is concluded if  $\frac{|Q_n(\bm{\theta}^{(t)})-Q_n(\bm{\theta}^{(t-1)})|}
{|Q_n(\bm{\theta}^{(t-1)})|}<10^{-4}.$

It is noted that Steps 2 and 3 are not standard CD algorithms, which iterate until convergence. Instead, only one iteration is taken, which can significantly reduce computational cost. Details on Steps 2.1 and 2.2 are provided in Appendix.
As the value of the objective function decreases at each step and is bounded below, the proposed algorithm is guaranteed to converge. Convergence is achieved in all of our numerical studies within 50 overall iterations.

\noindent\textbf{Tuning parameters} The proposed approach includes two tuning parameters $\lambda_1$ and $\lambda_2$, and one regularization parameter $r$. For $r$, published studies suggest setting it as fixed or examining a small number of values. We follow the literature \citep{Breheny09} and set $r=3$ in our numerical study. The values of $(\lambda_1,\lambda_2)$ are chosen using BIC.

\noindent\textbf{Parameter path} To better comprehend the proposed penalized estimation, we simulate one replicate under the linear model with MAF setting M1 and correlation structure AR(0.3). Details on the data settings are described in Section 3. With the proposed approach, we first examine the values of BIC as a function of $\lambda_1$ and $\lambda_2$ in Figure \ref{BIC_curve}. The optimal point with $(\lambda_1,\lambda_2)=(0.135,0.095)$ is clearly identified. We further examine the parameter paths in Figure \ref{para_path}. The proposed approach is observed to have parameter paths similar to those of other penalized estimates. The model is sparser with larger $\lambda_1$ and smoother with larger $\lambda_2$. For this simulated dataset, with the optimal tuning parameters, the proposed approach can correctly identify the majority of true positives while having a small number of false positives. More definitive results are presented below.

\noindent\textbf{Realization} To facilitate data analysis within and beyond this study, we have developed R code implementing the proposed approach and made it publicly available at www.github.com/shuanggema. The proposed approach is computationally affordable. For example, with fixed tuning parameters, for a simulated dataset with $q=5$, $p=5000$, and $n=250$, the analysis can be accomplished within one minute using a laptop with standard configurations.

\subsection{Statistical properties}
Consider the scenario where the number of G factors increases and the number of E factors is finite as the sample size increases. This reasonably fits the analyzed datasets and others.

Let $\bm{\theta}^0=\left(\left(\bm{\alpha}^0\right)',
\left(\bm{\beta}^0\right)',\left(\bm{\gamma}^0_1\right)',
\cdots,\left(\bm{\gamma}^0_q\right)'\right)'$ be the true parameter values, and \\ $\bm{\Theta}^0=\left(\left(\bm{\alpha}^0\right)',
\left(\bm{\beta}^0\right)',\left(\bm{\eta}^0_1\right)',
\cdots,\left(\bm{\eta}^0_q\right)'\right)'=\left(\left(\bm{\alpha}^0
\right)',\left(\bm{\beta}^0\right)',\left(\bm{\gamma}^0_1\odot
\bm{\beta}^0\right)',\cdots,\left(\bm{\gamma}^0_q\odot\bm{\beta}^0
\right)'\right)'$. Let $\mathcal{A}_1=\{j:\beta_j^0\neq 0\}$, $\mathcal{A}_2^k=\{j:\gamma_{kj}^0\neq0\textrm{ and }\beta^0_{j}\neq 0\}$, and $\mathcal{A}_2=\mathcal{A}_2^1\bigcup\cdots\bigcup \mathcal{A}_2^q$. Note that all $\alpha_k^0$'s are nonzero, and the corresponding parameters are not subject to penalization in estimation. With the hierarchical constraint, in $\mathcal{A}_2^k$, we are only interested in nonzero $\gamma_{kj}$'s for which the corresponding $\beta_j$'s are also nonzero. Here, we have $j\in \mathcal{A}_1$ if for some $k$, $j \in \mathcal{A}_2^k$. Denote $|\mathcal{A}|$ as the cardinality of set $\mathcal{A}$. Let $s=|\mathcal{A}_1|+|\mathcal{A}_2^1|+\cdots+|\mathcal{A}_2^q|$. For a vector $\bm{\nu}$ and index set $\mathcal{S}$, let $\bm{\nu}_{\mathcal{S}}$ denote the components of $\bm{\nu}$ indexed by $\mathcal{S}$. For a matrix $\bm{M$} and two index sets $\mathcal{S}_1$ and $\mathcal{S}_2$, denote $\bm{M}_{\mathcal{S}_1}$ and $\bm{M}_{\mathcal{S}_1\cdot}$ as the columns and rows of $\bm{M}$ indexed by $\mathcal{S}_1$, and $\bm{M}_{\mathcal{S}_1,\mathcal{S}_2}$ as the submatrix of $\bm{M}$ indexed by $\mathcal{S}_1$ and $\mathcal{S}_2$.

Denote $\bm{\theta}^{\ast}_{\mathcal{A}}=\left(\left(\bm{\alpha}^{\ast}
\right)',\left(\bm{\beta}^{\ast}_{\mathcal{A}_1}\right)',
\left(\bm{\gamma}^{\ast}_{1,\mathcal{A}_2^1}\right)',
\cdots,\left(\bm{\gamma}^{\ast}_{q,\mathcal{A}_2^q}\right)'\right)'$ as the minimizer of
\begin{eqnarray*}
\tilde{Q}_n(\bm{\theta}_{\mathcal{A}})&=&\frac{1}{2n}\left\| \bm{Y}-\bm{Z}\bm{\alpha}-\bm{X}_{\mathcal{A}_1}\bm{\beta}_{\mathcal{A}_1}
-\sum_{k=1}^q \bm{W}^{(k)}_{\mathcal{A}_2^k}\left(\bm{\beta}_{\mathcal{A}_2^k}\odot \bm{\gamma}_{k,\mathcal{A}_2^k}\right)\right\|_2^2\\
&&~~~~+\frac{1}{2}\lambda_2\left(\bm{\beta}_{\mathcal{A}_1}'
\bm{J}_{\mathcal{A}_1,\mathcal{A}_1}\bm{\beta}_{\mathcal{A}_1}+
\sum_{k=1}^q\bm{\gamma}_{k,\mathcal{A}_2^k}'\bm{J}_{\mathcal{A}_2^k,
\mathcal{A}_2^k}\bm{\gamma}_{k,\mathcal{A}_2^k}\right).
\end{eqnarray*}

The following conditions are assumed:

\begin{enumerate}
\item [\textbf{(C1)}] Components of the residual $\bm{\varepsilon}$ are i.i.d and sub-Gaussian with noise level $\sigma$. That is, for any vector $\bm{\nu}$ with $||\bm{\nu}||_2=1$ and any constant $\epsilon>0$, $P(|\bm{\nu}' \bm{\varepsilon}|\geq \epsilon)\leq 2 \exp\left(-\frac{\epsilon^2}{2\sigma^2}\right)$.

\item [\textbf{(C2)}] Let $b_0=\min\left\{\left\{|\beta^0_j|: j\in \mathcal{A}_1\right\},\left\{|\gamma_{kj}^0|: j\in \mathcal{A}_2^k, k=1,\cdots,q\right\}\right\}$. Then, $b_0 \sqrt{n/s}\rightarrow \infty$.

\item [\textbf{(C3)}]
Use $\lambda_{\min}(\bm{M})$ and $\lambda_{\max}(\bm{M})$ to denote the smallest and largest eigenvalues of matrix $\bm{M}$. Then,
\[\underset{\bm{\theta}_{\mathcal{A}}\in \mathcal{N}_0}{\max}\lambda_{\max}\left(\frac{1}{n}\bm{G}
(\bm{\beta}_{\mathcal{A}_2},\bm{\gamma}_{\mathcal{A}_1})'\bm{G}
(\bm{\beta}_{\mathcal{A}_2},\bm{\gamma}_{\mathcal{A}_1})\right)\leq s\bar{c},\]
and
\[\underset{\bm{\theta}_{\mathcal{A}}\in \mathcal{N}_0}{\min}\lambda_{\min}\left(\frac{1}{n}\bm{G}
(\bm{\beta}_{\mathcal{A}_2},\bm{\gamma}_{\mathcal{A}_1})'\bm{G}
(\bm{\beta}_{\mathcal{A}_2},\bm{\gamma}_{\mathcal{A}_1})+
\frac{1}{n}\bm{F}(\bm{\theta}_{\mathcal{A}})\right)\geq\underline{c},\]
where $\bm{\gamma}_{\mathcal{A}_1}=\left(\bm{\gamma}'_{1,\mathcal{A}_1},
\cdots,\bm{\gamma}'_{q,\mathcal{A}_1}\right)'$ with $\gamma_{kj}=0$, if $j\in \mathcal{A}_1$ but $j\notin \mathcal{A}_2^k$,
\[\bm{G}(\bm{\beta}_{\mathcal{A}_2},\bm{\gamma}_{\mathcal{A}_1})
=\left(\bm{Z}, \bm{U}(\bm{\gamma}_{\mathcal{A}_1}), \bm{V}^{(1)}(\bm{\beta}_{\mathcal{A}_2^1}),\bm{V}^{(2)}
(\bm{\beta}_{\mathcal{A}_2^2}),\cdots, \bm{V}^{(q)}(\bm{\beta}_{\mathcal{A}_2^q})\right)_{n\times (q+s)},
\]
with \[\bm{U}(\bm{\gamma}_{\mathcal{A}_1})=\bm{X}_{\mathcal{A}_1}+
\sum_{k=1}^q \bm{W}_{\mathcal{A}_1}^{(k)}\odot \left(\bm{1}_{n\times 1}\left(\bm{\gamma}_{k,{\mathcal{A}_1}}\right)'\right),~\bm{V}^{(k)}
(\bm{\beta}_{\mathcal{A}_2^k})=\bm{W}^{(k)}_{\mathcal{A}_2^k}\odot \left(\bm{1}_{n\times 1}\left(\bm{\beta}_{\mathcal{A}_2^k}\right)'\right),\]
$\bm{F}(\bm{\theta}_{\mathcal{A}})=\left(f_{jl}
(\bm{\theta}_{\mathcal{A}})\right)_{(q+s)\times (q+s)}$ with $f_{jl}(\bm{\theta}_{\mathcal{A}})=-
\left(\bm{W}_{\varsigma}^{(k)}\right)'(\bm{Y}-\bm{Z}\bm{\alpha}
-\bm{X}_{\mathcal{A}_1}\bm{\beta}_{\mathcal{A}_1}-\sum_{g=1}^q \bm{W}^{(g)}_{\mathcal{A}_2^{g}}(\bm{\beta}_{\mathcal{A}_2^{g}}\odot \bm{\gamma}_{g,\mathcal{A}_2^{g}}))$ if both $j$ and $l$ correspond to the $\varsigma$th element of $\mathcal{A}_2^k$, and 0 otherwise, $\mathcal{N}_0=\{\bm{\theta}_{\mathcal{A}}: ||\bm{\theta}_{\mathcal{A}}-\bm{\theta}^0_{\mathcal{A}}||_{\infty}\leq \frac{b_0}{2}\}$, and $\bar{c}$ and $\underline{c}$ are two positive constants.

\item [\textbf{(C4)}]$\lambda_2=O(\sqrt{1/n})$.
\item [\textbf{(C5)}] $\lambda_{\min}\left(\widetilde{\bm{J}}_{\mathcal{A},\mathcal{A}}
    \right)\geq0$ and $||\widetilde{\bm{J}}_{\mathcal{A},\mathcal{A}}
    \bm{\theta}^0_{\mathcal{A}}||_2=O(\sqrt{s})$, where $\widetilde{\bm{J}}_{\mathcal{A},\mathcal{A}}=\textrm{diag}\left(\bm{0}_{q\times q}, \bm{J}_{\mathcal{A}_1,\mathcal{A}_1}, \cdots, \bm{J}_{\mathcal{A}_2^q,\mathcal{A}_2^q}\right)$ is a block diagonal matrix with the diagonal blocks being $\bm{0}_{q\times q}, \bm{J}_{\mathcal{A}_1,\mathcal{A}_1}, \cdots,$ and $\bm{J}_{\mathcal{A}_2^q,\mathcal{A}_2^q}$.
\end{enumerate}

Condition (C1) is the sub-Gaussian condition which is commonly assumed in published studies \citep{Fan11,Guo16,Huang17}. Condition (C2) puts a lower bound on the size of the smallest signal. Condition (C3) assumes that the predictor matrix is ``well behaved''. Similar conditions have been assumed in \cite{Zou09}, \cite{Fan11}, and others. Condition (C4) restricts the rate of the tuning parameter $\lambda_2$. Condition (C5) makes a weak constraint on $\bm{J}$. It needs to be checked on a case-by-case basis, as $\bm{J}$ may vary across data. For the spline type penalty considered for SNP data, Condition (C5) is easily satisfied. For the Laplacian type penalty, it is also satisfied for example when the network is sparse.

\noindent \textbf{Theorem 1}: Under Conditions (C1)-(C5), there exists a local minimizer $\bm{\theta}^{\ast}_{\mathcal{A}}$ of $\tilde{Q}_n(\bm{\theta}_{\mathcal{A}})$ such that for any constant $E>0$,
\[P\left\{||\bm{\theta}^{\ast}_{\mathcal{A}}-
\bm{\theta}^0_{\mathcal{A}}||_2\leq \delta_n\right\}>1-\xi,\]
where $\delta_n=\frac{4\lambda_2||\widetilde{\bm{J}}_{\mathcal{A},
\mathcal{A}}\bm{\theta}^0_{\mathcal{A}}||_2}{\underline{c}}+E\sqrt{s/n}$ and $\xi=\exp\left(-\frac{\left[4\sqrt{n/s}\lambda_2||
\widetilde{\bm{J}}_{\mathcal{A},\mathcal{A}}
\bm{\theta}^0_{\mathcal{A}}||_2+E\underline{c}\right]^2}
{32\sigma^2\bar{c}}\right)$.

Proof is provided in Appendix. With Theorem 1, we have
\[||\bm{\theta}^{\ast}_{\mathcal{A}}-\bm{\theta}^0_{\mathcal{A}}||_2=
O_p(\sqrt{s/n})\textrm{ and }||\bm{\Theta}^{\ast}_{\mathcal{A}}-\bm{\Theta}^0_{\mathcal{A}}||_2=
O_p(\sqrt{s/n}),\]
as $\lambda_2=O(\sqrt{1/n})$ and $||\widetilde{\bm{J}}_{\mathcal{A},\mathcal{A}}
\bm{\theta}^0_{\mathcal{A}}||_2=O(\sqrt{s})$. This theorem establishes estimation consistency when the true sparsity structure is known.

Let $\mathcal{A}_1^c=\{j:\beta_j^0=0\}$ and $(\tilde{\mathcal{A}}_2^k)^c=\{j: \gamma^0_{kj}=0\textrm{ and }\beta^0_j\neq0\}$. Then we have $(\tilde{\mathcal{A}}_2^k)^c \bigcup \mathcal{A}_1^c=\{j: \eta^0_{kj}=0\}$. The following additional conditions are assumed.
\begin{enumerate}
\item [\textbf{(C6)}]
~~~~~~~~~~~~~$||\bm{U}(\bm{\gamma}^0_{\mathcal{A}_1^c})'\bm{G}
(\bm{\beta}^0_{\mathcal{A}_2},\bm{\gamma}^0_{\mathcal{A}_1})||_{2,\infty}
=O(n),$
$
\left|\left|\bm{V}^{(k)}\left(\bm{\beta}^0_{\left(\tilde{\mathcal{A}}_2^k
\right)^c}\right)'\bm{G}(\bm{\beta}^0_{\mathcal{A}_2},
\bm{\gamma}^0_{\mathcal{A}_1})\right|\right|_{2,\infty}=O(n),$
\[||\bm{U}(\bm{\gamma}^0_{j})||_2=O(\sqrt{n}), ||\bm{V}^{(k)}(\beta^0_j)||_2=O(\sqrt{n}), j=1,\cdots,p, \]
where for matrix $\bm{M}$, $||\bm{M}||_{2,\infty}=\max_{||\bm{\nu}||_2=1}||\bm{M\nu}||_{\infty}$, and $\bm{U}(\cdot)$ and $\bm{V}^{(k)}(\cdot)$ are defined in Condition (C3).
\[\max_{\bm{\theta}_{\mathcal{A}}\in \mathcal{N}_0}\max_j \lambda_{\max}\left(\bm{T}_1^{(j)}(\bm{\gamma}_{j})\right)=O(n),\]
where $\bm{T}_1^{(j)}(\bm{\gamma}_{j})=\left(t_{lh}^{(j)}(\bm{\gamma}_{j})\right)_{(q+s)\times (q+s)}$ with $t_{lh}^{(j)}(\bm{\gamma}_{j})=\left(\bm{X}_j+\sum\limits_{g=1}^q\bm{W}_j^{(g)}\gamma_{gj}\right)' \bm{W}_{\varsigma}^{(k)}$ if both $l$ and $h$ correspond to the $\varsigma$th element of $\mathcal{A}_2^k$, and 0 otherwise.
\[\max_{\bm{\theta}_{\mathcal{A}}\in \mathcal{N}_0}\max_j \lambda_{\max}\left(\bm{T}_2^{(j)}(\beta_j)\right)=O(n),\]
where $\bm{T}_2^{(j)}(\beta_j)=\left(t_{lh}^{(j)}(\beta_j)\right)_{(q+s)\times (q+s)}$ with $t_{lh}^{(j)}(\beta_j)=\left(\bm{W}_j^{(k)}\beta_j\right)' \bm{W}_{\varsigma}^{(k)}$ if both $l$ and $h$ correspond to the $\varsigma$th element of $\mathcal{A}_2^k$, and 0 otherwise.
\item [\textbf{(C7)}] $\log(p)=n^{a}, a\in(0,\frac{1}{2})$.
\item [\textbf{(C8)}] $\frac{\lambda_1}{\sqrt{s/n}}\rightarrow \infty,~ \frac{\lambda_1}{n^{a/2-1/2}\sqrt{\log{n}}}\rightarrow \infty$.
\item [\textbf{(C9)}] $b_0 \lambda_1^{-1}\rightarrow \infty$.
\end{enumerate}

Condition (C6) is similar to Condition 4 in \cite{Fan11}, where the first two equations control the ``correlations'' between the unimportant and important variables. Condition (C7) allows the number of G factors to increase as the sample size increases. Condition (C8) has also been assumed in \cite{Fan11} and others. Condition (C9) provides the rate at which the nonzero coefficients can be distinguished from zero \citep{Huang17}.

\noindent \textbf{Theorem 2}: Define $\hat{\bm{\theta}}$ as $\hat{\bm{\theta}}_{\mathcal{A}}=\bm{\theta}^{\ast}_{\mathcal{A}}$, $\hat{\bm{\beta}}_{\mathcal{A}_1^c}=0$, $\hat{\bm{\gamma}}_{k,(\tilde{\mathcal{A}}_2^k)^c}=0$ and $\hat{\bm{\gamma}}_{k,{\mathcal{A}_1^c}}$ being the minimizer of $Q_n(\bm{\theta})$ with other parameters fixed at the values defined as above. Then under Conditions (C1)-(C9), with probability tending to 1, $\hat{\bm{\theta}}$ is a strict local minimizer of  $Q_n(\bm{\theta})$.

Proof is provided in Appendix. With Theorem 2, we have $\hat{\bm{\eta}}_{k,\mathcal{A}_1^c}=0$ with $\hat{\bm{\beta}}_{\mathcal{A}_1^c}=0$, and $\hat{\bm{\eta}}_{k,(\tilde{\mathcal{A}}_2^k)^c}=0$ with $\hat{\bm{\gamma}}_{k,(\tilde{\mathcal{A}}_2^k)^c}=0$. Theorem 2 establishes the selection and estimation consistency properties of the proposed approach under high-dimensional settings.

\section{Simulation}
\begin{figure}[H]
  \centering
  \includegraphics[width=0.9\textwidth, angle=0]{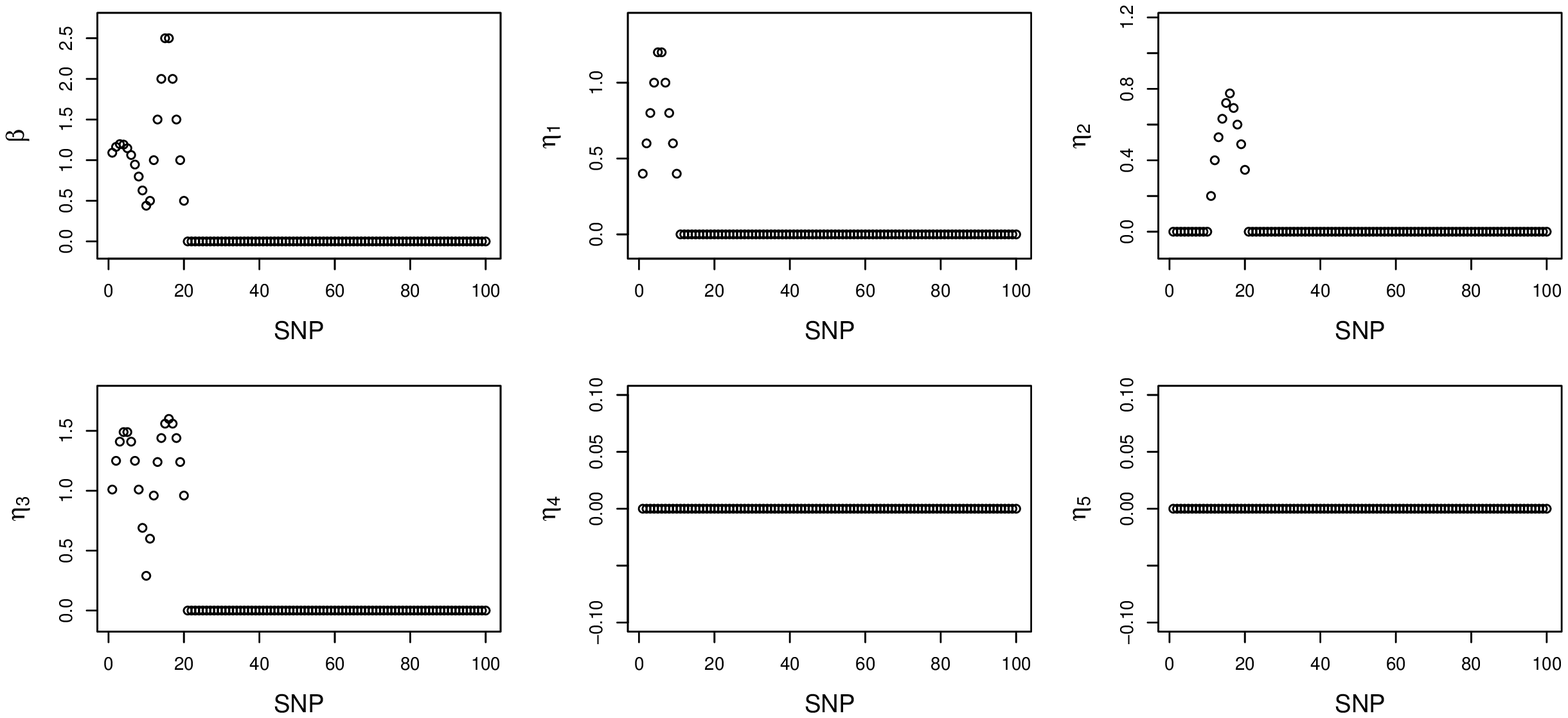}\\
  \caption{Simulation: true coefficient values for the main G effects and interactions. To improve presentation, only the first 100 effects are presented. The rest are zero.}\label{true_value}
\end{figure}

We simulate densely positioned SNP data, which have an adjacency structure. Specifically, (a) under all scenarios, $q=5$ and $p=5,000$. Thus, there are a total of 5,005 main effects and 25,000 interactions. (b) Two approaches, A1 and A2, are adopted to simulate G factors which mimic SNP data coded with three categories (0, 1, 2) for genotypes (aa, Aa, AA). (c) The A1 approach includes two steps, under which we first generate $p$ continuous variables from a multivariate Normal distribution with mean $\bm{0}$ and covariance matrix $\bm{\Sigma}=(\sigma_{jl})_{p\times p}$, and then dichotomize the continuous variables at $q_1$ and $q_2$ quantiles to generate 3-level G measurements (0, 1, 2). In the first step, two correlation structures are considered with different parameters. The first is the auto-regressive (AR) structure with $\sigma_{jl}=\rho^{|j-l|}$. We consider two levels of correlation with $\rho=0.3$ and $0.5$. The second is the banded correlation structure where two specific scenarios are considered. The first one (Band1) has $\sigma_{jl}=1$ if $j=l$, 0.3 if $|j-l|=1$, and 0 otherwise. The second one (Band2) has $\sigma_{jl}=1$ if $j=l$, 0.5 if $|j-l|=1$, 0.3 if $|j-l|=2$, and 0 otherwise. In the second step, the quantiles $q_1$ and $q_2$ are adjusted to generate G factors with different minor allele frequency (MAF) values. Consider two specific scenarios. Under the first scenario (M1), all of the G factors have MAF=0.05 with $q_1=0.91$ and $q_2=0.99$. Under the second one (M2), a half of the G factors have MAF=0.05, and the other half have MAF=0.15 with $q_1=0.73$ and $q_2=0.97$. (d) Under the A2 approach, we simulate G factors with the pairwise LD structure. Specifically, denote $p_A$ and $p_B$ as the MAFs of alleles A and B for two adjacent SNPs. The LD is defined as $\phi=r_{LD}\sqrt{p_A(1-p_A)p_B(1-p_B)}$ with pairwise correlation $r_{LD}$. Then, the four haplotypes ab, aB, Ab, AB have frequencies $(1-p_A)(1-p_B)+\phi$, $(1-p_A)p_B-\phi$, $p_A(1-p_B)-\phi$, and $p_Ap_B+\phi$, respectively. Following the literature \citep{Wu15}, with the Hardy-Weinberg equilibrium assumption, we simulate the SNP genotype (AA, Aa, aa) at locus 1 from a multinomial distribution given corresponding frequencies $(p_A^2,2p_A^2(1-p_A),(1-p_A)^2)$ and that at locus 2 accordingly from the conditional probability defined in \cite{Cui08}. Two pairwise correlations are considered with $r_{LD}=0.3$ and $r_{LD}=0.5$. For MAF, two scenarios similar to those in Step 2 of A1 are considered. (e) For E factors, we first generate five continuous variables from a multivariate Normal distribution with marginal mean 0, marginal variance 1, and AR correlation ($\rho=0.3$), and then dichotomize two of them at 0 and create two binary variables. There are thus three continuous and two binary E factors. (f) For E factors, their coefficients $\alpha_k, k=1,\cdots,5$ are generated from Uniform $(0.8, 1.2)$. There are 20 main G effects and 40 G-E interactions with nonzero coefficients. Two structures, the ``main effects, interactions'' hierarchial structure and the smoothness structure of SNP effects, are satisfied.
Specifically, we set $\beta_j=\sin(0.2j+0.9)+0.2$ for $j=1,\cdots,10$, $\beta_j=0.5(j-10)$ for $j=11,\cdots,15$, $\beta_j=0.5(21-j)$ for $j=16,\cdots,20$, $\eta_{1j}=0.2j+0.2$ for $j=1,\cdots,5$, $\eta_{1j}=0.2(11-j)+0.2$ for $j=6,\cdots,11$, $\eta_{2j}=0.2\sqrt{3j-32}$ for $j=11,\cdots,15$, $\eta_{2j}=0.2\sqrt{63-3j}$ for $j=16,\cdots,20$, $\eta_{3j}=-(0.2j-0.9)^2+1.5$ for $j=1,\cdots,10$, and $\eta_{3j}=-(0.2j-3.2)^2+1.6$ for $j=11,\cdots,20$. The rest of the effects are zero. A graphical presentation is provided in Figure \ref{true_value}, where the sparsity and smoothness of effects are easy to see. (g) Consider two types of response variables and models. The first is a continuous response under model (\ref{general_model}). The second is a censored survival response under the AFT model, where the censoring times are generated from an exponential distribution with parameter adjusted to achieve $\sim 20\%$ censoring. For both models, the random error $\varepsilon_i$ follows a standard Normal distribution. (h) Set the sample size $n=250$ and $n=350$ for the continuous and survival settings, respectively. There are a total of 24 scenarios, comprehensively covering a wide spectrum with different types of responses and correlation structures among G factors, and various levels of MAF.

For the simulated data, we consider the proposed approach with the spline type penalty defined in (\ref{spline}). We also consider the following alternatives. \textbf{MA}, which is a marginal analysis approach that analyzes one G factor along with all E factors and corresponding interactions at a time. P-values of the G factors and interactions are adjusted using the false discovery rate (FDR) approach. This approach has been commonly adopted in published studies and is a suitable benchmark for comparison. \textbf{HierMCP}, which is the non-structured counterpart of the proposed approach, where the MCP penalty is applied for estimation and selection. Comparing with this approach can reveal the value of incorporating the two structures.
\textbf{SMCP}, which is based on model (\ref{general_model}) and imposes the MCP and structured penalties on $\beta_j$ and $\eta_{kj}$ without respecting the ``main effects, interactions'' hierarchy. Comparing with this approach can reveal the value of the special consideration on interactions. We acknowledge that there are other interaction analysis approaches that can be applied to the simulated data. The above alternatives are adopted as they are perhaps the most relevant. Comparing with them can in a relatively direct way establish the merit of the proposed structured penalization and decomposition strategy for interaction analysis.

When evaluating identification performance, both main effects and interactions are considered. Measures used include the number of true positives (M:TP) and false positives (M:FP) for main effects, and number of true positives (I:TP) and false positives (I:FP) for interactions. Estimation performance is assessed using the root sum of squared errors (RSSE) defined as $||\hat{\bm{\Theta}}-\bm{\Theta}^0||_2$, where $\hat{\bm{\Theta}}$ and $\bm{\Theta}^0$ are the estimated and true values of $\bm{\Theta}=(\bm{\alpha}',\bm{\beta}',\bm{\eta}_1',\cdots,
\bm{\eta}_q')'$. We also take the underlying structure of SNPs into consideration and compute the root structured error (RSE) $\sqrt{(\hat{\bm{\Theta}}-\bm{\Theta}^0)'\tilde{\bm{J}}
(\hat{\bm{\Theta}}-\bm{\Theta}^0)}$, where $\widetilde{\bm{J}}=\textrm{diag}\left(\bm{0}_{q\times q}, \bm{J},\cdots, \bm{J}\right)$. For evaluating prediction performance, an independent testing set with 100 subjects is generated for each simulated dataset. We adopt the prediction mean squared error (PMSE) for continuous outcomes and C-statistic (Cstat) for censored survival outcomes. C-statistic is the time-integrated area under the time-dependent ROC framework and measures the overall adequacy of risk prediction for censored survival data, with a larger value indicating better prediction.

For each scenario, 500 replicates are simulated, and the means and standard deviations (sd) of the evaluation measures are computed. Summary results under the linear model with MAF settings M1 and M2 are shown in Tables \ref{Scenario1-6} and \ref{Scenario7-12}, respectively. The rest of the results are shown in Appendix. Across all simulation scenarios, the proposed approach is observed to have superior or similar performance compared to the alternatives. Specifically, it can more accurately identify both the true main effects and interactions while having a small number of false positives. For example in Table \ref{Scenario1-6} with AR(0.3), the proposed approach has (M:TP,M:FP,I:TP,I:FP)=(19.7,0.0,33.8,4.1), compared to (0.1,11.2,2.2,77.9) for MA, (11.7,68.5,3.4,4.2) for HierMCP, and (17.4,2.7,23.4,19.7) for SMCP. Compared to MA and HierMCP, the proposed approach has much better identification performance, which provides a strong support to the structured analysis strategy. It also outperforms SMCP, which suggests the effectiveness of the proposed decomposition strategy for respecting the interaction hierarchy. The advantage of the proposed approach gets more prominent under MAF setting M2. For example in Table \ref{Scenario7-12} with Band1, the proposed approach has (M:TP,M:FP,I:TP,I:FP)=(19.7,1.0,33.3,5.1), compared to (0.1,6.7,1.6,53.6) for MA, (11.7,64.7,3.9,5.2) for HierMCP, and (16.1,7.0,11.3,74.1) for SMCP. We also observe the superiority of the proposed approach in estimation. For example in Table \ref{Scenario1-6} with LD(0.5), the proposed approach has RSSE=2.95, compared to 16.15 (MA), 17.76 (HierMCP), and 4.93 (SMCP). It also has smaller structured errors. In addition, the proposed approach has satisfactory prediction performance. For example in Table \ref{Scenario7-12} with Band2, the PMSEs are 29.94 (MA), 23.04 (HierMCP), 4.18 (SMCP), and 1.59 (proposed). The observed patterns for data with survival outcomes (Tables \ref{Scenario13-18} and \ref{Scenario19-24}) are similar, where the proposed approach performs better than or comparable to the alternatives.

For SNP data, we have also examined a few other simulation scenarios, and the observed patterns are similar (details omitted). We have also experimented with continuously distributed G measurements, which mimic gene expression data, and applied the Laplacian type penalty function. Similar superiority of the proposed approach is observed (details omitted).

\section{Data analysis}

\subsection{GENEVA diabetes data (NHS/HPFS)}

The Gene Environment Association Studies (GENEVA) consortium is part of the Genes, Environment and Health Initiative (GEI) organized by the NIH. We analyze the GENEVA Type 2 Diabetes data, where the goal is to identify genetic factors that are associated with type 2 diabetes phenotypes, biomarkers, and others. In our analysis, data are downloaded from dbGaP (accession number phs000091.v2.p1). The response variable of interest is body mass index (BMI), which is continuously distributed. BMI level is one of the most important risk factors for type 2 diabetes. Following recent published studies, we take a ``loose'' definition of E factors. Specifically, E factors considered include age, family history of diabetes among first degree relatives (famdb), total physical activity (act), trans fat intake (trans), cereal fiber intake (ceraf), and heme iron intake (heme), all of which have been suggested to be potentially associated with BMI and diabetes. For G factors, we analyze SNPs on chromosome 4, which plays an important role in many disorders, such as Parkinson's disease, Huntington's disease, and others. Preprocessing similar to that in \cite{Wu14} is conducted, which includes subject matching, standard quality control for SNPs, and missing data imputation. Data are available on 2,558 subjects and 40,568 SNPs. As the number of relevant SNPs is not expected to be large, to improve stability, we conduct a marginal screening. Specifically, a p-value is computed for each SNP based on a marginal linear model. Then the region of 10,000 consecutive SNPs with the smallest sum of p-values is selected for downstream analysis. With the physical adjacency structure in mind, in the prescreening, we select a region (as opposed to individual SNPs).

We adopt the linear regression model and spline type penalty (\ref{spline}). The proposed approach identifies 71 main SNP effects and 128 G-E interactions. The detailed estimation results are provided in Table \ref{Tab:GENEVA} and also presented in Figure \ref{Fig:GENEVA}, where SNPs are sorted according to their physical locations on the chromosome. In terms of main effects, three E factors, age, act, and ceraf, have negative coefficients, and the other three, famdb, trans, and heme, have positive coefficients, which are consistent with findings in the literature. Figure \ref{Fig:GENEVA} shows that the estimated effects demonstrate a certain degree of smoothness, which fits the design of the proposed approach. Genes that the identified SNPs belong to or are the closest to are also provided in Table \ref{Tab:GENEVA}. Literature search suggests that these genes and interactions may have important implications, which may provide support to the validity of the proposed approach. For example, gene NPFFR2 has been found to play an important role in obesity predisposition, and some NPFFR2 haplotypes have been suggested to be strongly protective against obesity. Gene CXCL2 has been shown to be up-regulated in obese subjects and contribute to the chemotaxis of neutrophils which are one type of circulating cells greatly activated in obese subjects. Published analysis has also found that the enzyme encoded by gene GK2 plays a key role in the regulation of glycerol uptake and metabolism, and its activity in human adipose tissue is related to obesity.

Beyond the proposed approach, we also conduct analysis using the alternatives. The summary comparison results are shown in Table \ref{Tab:overlap}. It is observed that the proposed approach identifies different main G effects and more significantly different interactions from those with the alternatives. Without reinforcing the interaction hierarchical structure, SMCP identifies the smallest number of main effects but the second largest number of interactions. Both the proposed approach and HierMCP identify a moderate number of main effects and interactions.

With real data, it is difficult to objectively evaluate identification accuracy. To provide support to the identification results, we examine prediction performance and selection stability using a resampling-based approach  \citep{Huang10}. With 500 resamplings, we compute the mean PMSEs, which are 15.38 (MA), 17.47 (HierMCP), 13.11 (SMCP), and 13.06 (proposed). The proposed approach has prediction performance comparable to SMCP and better than MA and HierMCP. We further compute the observed occurrence index (OOI) to measure selection stability. It is the probability of a specific main effect or interaction identified in 500 resamplings. The mean OOI values for the identified main G effects and interactions using the proposed approach is 0.69, compared to 0.47 (MA), 0.39 (HierMCP), and 0.21 (SMCP). The proposed approach has a prominent superiority in selection stability.

\subsection{TCGA skin cutaneous melanoma data}
We consider The Cancer Genome Atlas (TCGA) skin cutaneous melanoma (SKCM) data. TCGA is a collective effort organized by NIH and has published high quality clinical, environmental, and genetic data. We focus on the processed level 3 data, which are downloaded from TCGA Provisional using the R package \textit{cgdsr}. As in several recent published studies, we analyze the (censored) overall survival. The analyzed E factors include age, AJCC nodes pathologic stage (PN), gender, Breslow's depth, and Clark level, all of which have been extensively studied in the literature. For G factors, we consider the mRNA gene expressions. In TCGA, gene expression measurements are the z-scores, which have been lowess-normalized, log-transformed and median-centered, and quantify the relative expressions of tumor samples with respect to normal. Data are available on 298 subjects and 18,934 gene expressions. Among the subjects, 152 died during followup. Marginal screening is also conducted, and the 10,000 genes with the smallest p-values are selected for downstream analysis. Here as genes can be physically far from each other, the screening is directly based on p-values to select individual genes.

With a censored survival outcome, we adopt the AFT model. Examining the estimation procedure described in Appendix suggests that the proposed computational algorithm can be directly applied. With gene expression measurements, we adopt the Laplacian type penalty (\ref{Laplacian}). The proposed analysis identifies 50 main G effects and 44 interactions. The detailed estimation results are provided in Table \ref{Tab:SKCM}. All five E factors except for gender have negative coefficients, which match observations in the literature. The identified genes are also presented in Figure \ref{Fig:SKCM}, where two genes are connected if they have a nonzero adjacency value. For the identified genes, published studies provide independent evidences of their associations with cutaneous melanoma. For example, ACTL6A (BAF53) is a subunit of the SWI/SNF complex which has been found to be critical for the expression of microphthalmia-associated transcription factor in melanoma cells. FAM131B-BRAF fusion has been observed to comprise an alternative mechanism of MAPK pathway activation, and MAPK pathway plays important roles in melanoma etiology, prognosis, and treatment. Gene GOLPH3 has been shown to regulate cell size and enhance growth-factor-induced mTOR signaling in melanoma cells, and suggested as a new oncogene that is commonly targeted for amplification in melanoma. Gene IL17A has been found to have tumorigenic effects in melanoma cell lines, which are related to the signal transducer and activator of transcription pathway signaling. It has been demonstrated that mutations in RAC1 are potentially biologically associated with cutaneous melanoma, and the pharmacological inhibition of downstream effectors of RAC1 signaling could be of therapeutic benefit. In addition, gene SERPINB3 has been reported to be up-regulated in benign hyperplasia in melanoma.

Analysis is further conducted using the three alternatives, and the summary comparison results are presented in Table \ref{Tab:overlap}. As for the previous dataset, the proposed approach identifies different sets of main effects and interactions. We also evaluate prediction performance and selection stability. In prediction evaluation, the mean C-statistics over 500 resamplings are 0.54 (MA), 0.59 (HierMCP), 0.64 (SMCP), and 0.65 (Proposed). In addition, the average OOI of the proposed approach is 0.87, compared to 0.53 (MA), 0.55 (HierMCP), and 0.77 (SMCP). The proposed approach again has better prediction performance and stability.

\section{Discussion}

For G-E interaction analysis, in this article, we have developed a new approach which shares similar desirable properties as the existing ones but also advances from them by accommodating the underlying structures of G factors. Although structured analysis has been conducted for main G effects in some recent publications, this study is among the first to conduct structured analysis in the context of G-E interaction analysis. Significant complexity is brought by the multiple effects (coefficients) that correspond to one G factor and the need to respect the ``main effects, interactions'' hierarchy. The proposed approach belongs to the well-established penalization paradigm and has an intuitive definition. Although it has multiple penalty terms, it is computationally much manageable. It is proved to have the consistency properties, which have not been established for most alternatives and provide a uniquely strong ground for the proposed approach. Extensive numerical studies show the practical superiority. Overall, this study provides a practically useful new way for analyzing G-E interactions.

Although described using the linear regression model for a continuous response as an example, the proposed approach can be extended to other data settings/models. It can accommodate multiple types of structures, as long as the $\bm{J}$ matrix satisfies certain mild conditions. As shown in published studies, $\bm{J}$ may need to be defined on a case-by-case basis. We leave it to future research to study the definition and properties of $\bm{J}$ for other types of omics data.

\clearpage
\section*{Acknowledgments}
This work was partly supported by the National Institutes of Health [CA216017, CA204120]; National Natural Science Foundation of China [61402276, 91546202]; and Bureau of Statistics of China [2016LD01, 2018LD02].

%\clearpage

\clearpage

\begin{center}
\begin{table}[h]
\caption{Simulation results under the linear model with MAF setting M1. In each cell, mean (sd) based on 500 replicates.} \label{Scenario1-6}
\centering
\renewcommand{\tabcolsep}{0.2pc} % enlarge column spacing
\begin{tabular}{llllllll}
\hline
 & M:TP  &  M:FP  & I:TP  & I:FP  & RSSE  & RSE  & PMSE \\
\hline
\multicolumn{8}{c}{ AR(0.3) }\\
MA & 0.1(0.5) & 11.2(15.9) & 2.2(2.2) & 77.9(79.0) & 15.03(5.07) & 30.32(17.12) & 28.36(8.80) \\
HierMCP & 11.7(1.7) & 68.5(11.4) & 3.4(1.6) & 4.2(2.0) & 13.29(1.04) & 26.48(2.53) & 20.45(4.49) \\
SMCP & 17.4(4.1) & 2.7(5.2) & 23.4(4.8) & 19.7(14.3) & 5.35(0.94) & 2.65(0.67) & 2.05(0.39) \\
Proposed & 19.7(0.7) & 0.0(0.1) & 33.8(3.3) & 4.1(2.5) & 3.09(0.82) & 2.32(0.66) & 1.47(0.31) \\
\hline
\multicolumn{8}{c}{ AR(0.5) }\\
MA & 0.4(1.0) & 15.0(18.0) & 4.8(3.7) & 106.9(79.7) & 20.15(8.82) & 44.63(27.13) & 40.81(16.31) \\
HierMCP & 12.5(1.5) & 78.1(13.7) & 4.1(1.8) & 5.3(2.5) & 14.54(1.30) & 30.51(3.10) & 25.42(6.31) \\
SMCP & 19.0(1.4) & 3.0(5.2) & 23.6(4.7) & 20.4(16.5) & 5.15(0.88) & 2.71(0.67) & 2.28(0.70) \\
Proposed & 19.7(0.6) & 0.0(0.3) & 34.8(2.8) & 3.1(2.0) & 2.67(0.75) & 2.39(0.69) & 1.47(0.37) \\
\hline
\multicolumn{8}{c}{ Band1 }\\
MA & 0.2(0.8) & 10.4(17.0) & 1.9(2.3) & 75.7(81.1) & 13.42(3.38) & 24.02(13.53) & 24.63(6.17) \\
HierMCP & 11.6(1.6) & 70.3(10.5) & 3.0(1.8) & 4.0(2.0) & 13.36(0.97) & 26.30(2.41) & 20.91(4.06) \\
SMCP & 17.7(3.1) & 3.5(5.8) & 22.0(4.2) & 20.8(15.3) & 5.48(0.92) & 2.71(0.60) & 2.19(0.55) \\
Proposed & 19.6(0.8) & 0.0(0.4) & 33.4(3.5) & 4.3(2.9) & 3.24(0.99) & 2.40(0.72) & 1.55(0.40) \\
\hline
\multicolumn{8}{c}{ Band2 }\\
MA & 0.2(0.5) & 9.2(14.6) & 3.1(3.1) & 79.2(80.3) & 15.09(4.99) & 29.89(17.11) & 34.68(10.47) \\
HierMCP & 12.4(1.7) & 76.1(14.2) & 3.9(1.9) & 5.4(2.9) & 14.22(1.39) & 29.54(3.48) & 24.11(6.01) \\
SMCP & 18.8(1.7) & 2.2(3.8) & 24.4(4.8) & 18.8(14.1) & 4.93(1.00) & 2.72(0.60) & 2.17(0.53) \\
Proposed & 19.6(0.6) & 0.0(0.0) & 34.2(3.6) & 3.4(2.2) & 2.74(0.92) & 2.40(0.77) & 1.49(0.41) \\
\hline
\multicolumn{8}{c}{ LD(0.3) }\\
MA & 0.2(0.7) & 8.5(13.8) & 3.0(2.8) & 70.6(75.7) & 14.40(4.10) & 27.57(13.99) & 27.24(7.68) \\
HierMCP & 11.9(1.7) & 93.7(10.8) & 1.6(1.2) & 1.6(1.3) & 15.52(1.15) & 32.24(2.91) & 25.96(5.44) \\
SMCP & 17.3(4.1) & 3.0(4.7) & 22.9(4.7) & 15.4(12.1) & 5.42(0.97) & 2.68(0.59) & 2.23(0.65) \\
Proposed & 19.3(1.0) & 0.0(0.1) & 33.2(3.8) & 3.5(2.6) & 3.10(0.98) & 2.44(0.73) & 1.60(0.44) \\
\hline
\multicolumn{8}{c}{ LD(0.5) }\\
MA & 0.4(1.1) & 9.5(16.3) & 5.0(3.9) & 77.8(73.9) & 16.15(5.37) & 34.21(16.84) & 33.86(10.10) \\
HierMCP & 12.3(1.6) & 109.5(14.8) & 1.6(1.1) & 2.1(1.4) & 17.76(1.62) & 38.96(4.07) & 35.11(9.11) \\
SMCP & 18.6(2.3) & 2.4(3.6) & 25.3(4.9) & 15.7(14.0) & 4.93(1.16) & 2.61(0.59) & 2.20(0.62) \\
Proposed & 19.2(1.1) & 0.1(0.4) & 33.7(3.8) & 2.7(2.6) & 2.95(1.10) & 2.60(0.89) & 1.60(0.50) \\
\hline
\end{tabular}
\end{table}
\end{center}

\begin{center}
\begin{table}[h]
\caption{Simulation results under the linear model with MAF setting M2. In each cell, mean (sd) based on 500 replicates.} \label{Scenario7-12}
\centering
\renewcommand{\tabcolsep}{0.2pc} % enlarge column spacing
\begin{tabular}{llllllll}
\hline
 & M:TP  &  M:FP  & I:TP  & I:FP  & RSSE  & RSE  & PMSE \\
\hline
\multicolumn{8}{c}{ AR(0.3) }\\
MA & 0.1(0.5) & 7.1(14.4) & 2.0(2.1) & 53.9(70.5) & 11.20(1.62) & 17.58(8.90) & 23.30(5.04) \\
HierMCP & 11.9(1.7) & 64.4(11.0) & 4.2(2.1) & 5.7(2.3) & 13.09(1.00) & 26.28(2.37) & 19.38(4.95) \\
SMCP & 16.5(3.3) & 6.5(9.9) & 12.3(8.3) & 68.7(25.9) & 7.06(1.51) & 3.56(1.05) & 5.53(3.43) \\
Proposed & 19.7(0.6) & 0.0(0.1) & 34.2(3.3) & 4.0(2.2) & 3.04(0.86) & 2.26(0.53) & 1.45(0.29) \\
\hline
\multicolumn{8}{c}{ AR(0.5) }\\
MA & 0.3(0.9) & 10.3(15.7) & 4.0(3.6) & 80.0(79.4) & 14.89(4.30) & 30.05(15.14) & 36.06(12.01) \\
HierMCP & 12.5(1.4) & 70.2(14.1) & 5.0(2.4) & 7.3(3.5) & 14.02(1.58) & 29.43(3.89) & 23.00(6.29) \\
SMCP & 17.7(3.1) & 4.9(8.1) & 17.8(5.9) & 54.8(26.8) & 6.10(1.12) & 3.06(0.83) & 4.09(3.23) \\
Proposed & 19.7(0.6) & 0.4(2.8) & 34.7(2.9) & 3.5(2.5) & 2.72(0.77) & 2.45(0.78) & 1.50(0.40) \\
\hline
\multicolumn{8}{c}{ Band1 }\\
MA & 0.1(0.8) & 6.7(13.3) & 1.6(2.1) & 53.6(69.2) & 10.56(1.14) & 14.71(8.46) & 22.79(4.91) \\
HierMCP & 11.7(1.5) & 64.7(10.1) & 3.9(2.2) & 5.2(2.7) & 13.12(0.96) & 25.96(2.44) & 19.76(4.25) \\
SMCP & 16.1(3.4) & 7.0(10.8) & 11.3(7.7) & 74.1(23.7) & 7.19(1.35) & 3.59(0.92) & 5.95(3.14) \\
Proposed & 19.7(0.8) & 1.0(4.3) & 33.3(3.5) & 5.1(4.7) & 3.24(1.02) & 2.51(0.83) & 1.58(0.45) \\
\hline
\multicolumn{8}{c}{ Band2 }\\
MA & 0.1(0.5) & 6.2(13.2) & 2.6(2.8) & 55.1(70.7) & 11.86(2.08) & 19.93(10.16) & 29.94(7.08) \\
HierMCP & 12.6(1.6) & 69.6(17.0) & 4.9(2.4) & 6.8(2.9) & 13.84(1.62) & 28.73(3.97) & 23.04(6.61) \\
SMCP & 16.9(3.8) & 5.2(8.8) & 17.8(7.2) & 59.3(24.8) & 6.18(1.22) & 3.09(0.83) & 4.18(2.65) \\
Proposed & 19.6(0.7) & 1.2(5.7) & 33.8(3.6) & 4.1(4.0) & 2.82(1.02) & 2.55(0.92) & 1.59(0.57) \\
\hline
\multicolumn{8}{c}{ LD(0.3) }\\
MA & 0.2(0.7) & 8.5(13.8) & 3.0(2.8) & 70.6(75.7) & 14.40(4.10) & 27.57(13.99) & 27.24(7.68) \\
HierMCP & 11.9(1.7) & 93.7(10.8) & 1.6(1.2) & 1.6(1.3) & 15.52(1.15) & 32.24(2.91) & 25.96(5.44) \\
SMCP & 17.3(4.1) & 3.0(4.7) & 22.9(4.7) & 15.2(12.2) & 5.42(0.97) & 2.67(0.59) & 2.22(0.65) \\
Proposed & 19.3(1.0) & 0.0(0.1) & 33.2(3.8) & 3.5(2.6) & 3.10(0.98) & 2.44(0.73) & 1.60(0.44) \\
\hline
\multicolumn{8}{c}{ LD(0.5) }\\
MA & 0.4(1.1) & 9.5(16.3) & 5.0(3.9) & 77.8(73.9) & 16.15(5.37) & 34.21(16.84) & 33.86(10.10) \\
HierMCP & 12.3(1.6) & 109.5(14.8) & 1.6(1.1) & 2.1(1.4) & 17.76(1.62) & 38.96(4.07) & 35.11(9.11) \\
SMCP & 18.5(2.3) & 2.3(3.5) & 25.2(4.9) & 15.5(14.0) & 4.93(1.17) & 2.60(0.59) & 2.19(0.62) \\
Proposed & 19.2(1.1) & 0.1(0.4) & 33.7(3.8) & 2.7(2.6) & 2.95(1.10) & 2.60(0.89) & 1.60(0.50) \\
\hline
\end{tabular}
\end{table}
\end{center}

\begin{center}
\begin{table}
\caption{Analysis of the GENEVA diabetes data (NHS/HPFS) using the proposed approach: identified main effects and interactions.} \label{Tab:GENEVA}
\renewcommand{\tabcolsep}{0.25pc} % enlarge column spacing
\centering
\begin{tabular}{lclccccccc}
\hline
~~~~~SNP	&	Location	&	~~~~~Gene$^*$	&		&	age	&	famdb	&	act	&	trans	&	ceraf	&	heme	\\
\hline
		&		&		&		&	-0.3331	&	0.1711	&	-0.2659	&	0.2185	&	-0.3332	&	0.5615	\\
rs17090278	&	61679934	&	RP11-593F5.2	&	-0.0016	&		&		&		&		&		&		\\
rs10019557	&	61684734	&	RP11-593F5.2	&	-0.0019	&		&		&		&		&		&		\\
rs17090285	&	61695580	&	RP11-593F5.2	&	-0.0019	&		&		&		&		&		&		\\
rs17090286	&	61695978	&	RP11-593F5.2	&	-0.0016	&		&		&		&		&		&		\\
rs11731112	&	65554491	&	RP11-63H19.1	&	0.0015	&		&		&		&		&		&		\\
rs4355422	&	65557183	&	RP11-63H19.1	&	0.0015	&		&		&		&		&		&		\\
rs1430504	&	65681190	&	RP11-707A18.1	&	-0.0035	&		&		&		&		&		&		\\
rs6551878	&	65690589	&	RP11-707A18.1	&	-0.0056	&		&		&		&		&		&		\\
rs6823601	&	65691562	&	RP11-707A18.1	&	-0.0043	&		&		&		&		&		&		\\
rs13151560	&	67160442	&	MIR1269A	&	0.0025	&		&		&		&		&		&		\\
rs1858306	&	67161812	&	MIR1269A	&	0.0083	&	0.0021	&	0.002	&	-0.0013	&	0.0027	&		&		\\
rs10016795	&	67167064	&	MIR1269A	&	0.0174	&	0.0047	&	0.0074	&	-0.0054	&	0.0096	&	-0.0026	&	-0.0034	\\
rs17087008	&	67188696	&	MIR1269A	&	0.0282	&	0.0049	&	0.0169	&	-0.0145	&	0.0213	&	-0.0046	&	-0.0119	\\
rs12331987	&	67188980	&	MIR1269A	&	0.0373	&		&	0.0261	&	-0.0265	&	0.0308	&	-0.0056	&	-0.0256	\\
rs10000219	&	67200024	&	MIR1269A	&	0.0405	&	-0.0105	&	0.0274	&	-0.0336	&	0.0304	&	-0.0028	&	-0.0357	\\
rs4860208	&	67201368	&	MIR1269A	&	0.035	&	-0.0142	&	0.02	&	-0.0277	&	0.0197	&		&	-0.0301	\\
rs1511286	&	67213473	&	MIR1269A	&	0.0215	&	-0.0077	&	0.0082	&	-0.0126	&	0.0072	&		&	-0.0139	\\
rs1033095	&	67232011	&	RPS23P3	&	0.0064	&	-0.0012	&	0.0011	&	-0.0019	&		&		&	-0.002	\\
rs11936928	&	67489994	&	RPS23P3	&	0.0014	&		&		&		&		&		&		\\
rs6838523	&	67494918	&	RPS23P3	&	0.0014	&		&		&		&		&		&		\\
rs10033058	&	69177408	&	YTHDC1	&	0.0055	&		&		&		&		&		&		\\
rs2293595	&	69178920	&	YTHDC1	&	0.0097	&		&		&		&		&		&	0.0031	\\
rs12649108	&	69181942	&	YTHDC1	&	0.0095	&		&		&		&		&		&	0.0036	\\
rs17089267	&	69183791	&	YTHDC1	&	0.0067	&		&		&		&		&		&	0.002	\\
rs1730872	&	69189048	&	YTHDC1	&	0.0018	&		&		&		&		&		&		\\
rs1399247	&	70973970	&	CSN1S2AP	&	-0.0012	&		&		&		&		&		&		\\
rs1717600	&	70974315	&	CSN1S2AP	&	-0.0013	&		&		&		&		&		&		\\
rs11936367	&	72884978	&	NPFFR2	&	-0.0013	&		&		&		&		&		&		\\
rs7699403	&	72893324	&	NPFFR2	&	-0.0042	&		&		&		&		&		&		\\
rs6856651	&	72896457	&	NPFFR2	&	-0.0068	&	0.0013	&		&		&		&		&	-0.001	\\
rs7654531	&	72900621	&	NPFFR2	&	-0.0079	&	0.0018	&		&		&		&		&	-0.0013	\\
rs6824342	&	72903182	&	NPFFR2	&	-0.0074	&	0.0016	&		&		&		&		&		\\
rs6824703	&	72903318	&	NPFFR2	&	-0.0051	&		&		&		&		&		&		\\
rs7687603	&	72915996	&	NPFFR2	&	-0.002	&		&		&		&		&		&		\\
rs12649753	&	74940765	&	CXCL2	&	0.0092	&		&		&	-0.0023	&		&		&	0.0026	\\
rs546829	&	74956372	&	CXCL2	&	0.0199	&	-0.0012	&	-0.0028	&	-0.012	&	0.0024	&	0.0027	&	0.0129	\\
rs1837559	&	74959093	&	CXCL2	&	0.0257	&	-0.003	&	-0.0061	&	-0.0217	&	0.0033	&	0.0034	&	0.0226	\\
\hline \multicolumn{10}{r}{{Continued on the next page}}
\end{tabular}
\end{table}
\end{center}

\setcounter{table}{2}

\begin{center}
\begin{table}
\caption{Continued from the previous page} \label{Tab:GENEVA}
\renewcommand{\tabcolsep}{0.2pc} % enlarge column spacing
\centering
\begin{tabular}{lclccccccc}
\hline
~~~~~SNP	&	Position	&	~~~~~Gene	&		&	age	&	famdb	&	act	&	trans	&	ceraf	&	heme	\\
\hline
rs9131	&	74963049	&	CXCL2	&	0.0232	&	-0.0038	&	-0.0066	&	-0.0196	&	0.0021	&	0.0011	&	0.0199	\\
rs1866755	&	74978340	&	MTHFD2L	&	0.0156	&	-0.0024	&	-0.0038	&	-0.0098	&		&		&	0.0096	\\
rs7686861	&	74998484	&	MTHFD2L	&	0.0064	&		&		&	-0.0019	&		&		&	0.0018	\\
rs11737437	&	80262521	&	NAA11	&	-0.0019	&		&		&		&		&		&		\\
rs10004440	&	80272792	&	NAA11	&	-0.0043	&		&		&		&		&		&		\\
rs2903619	&	80281513	&	NAA11	&	-0.0056	&		&		&		&		&		&		\\
rs11731223	&	80290084	&	GK2	&	-0.0051	&		&		&		&		&		&		\\
rs6534350	&	80305179	&	GK2	&	-0.0025	&		&		&		&		&		&		\\
rs17003746	&	80314643	&	GK2	&	-0.0015	&		&		&		&		&		&		\\
rs11930550	&	80317724	&	GK2	&	-0.0019	&		&		&		&		&		&		\\
rs17003749	&	80317772	&	GK2	&	-0.0014	&		&		&		&		&		&		\\
rs7680648	&	82666782	&	RP11-689K5.3	&	-0.0018	&		&		&		&		&		&		\\
rs17561568	&	82667783	&	RP11-689K5.3	&	-0.0068	&	-0.0011	&		&		&		&		&		\\
rs35036928	&	82671170	&	RP11-689K5.3	&	-0.0137	&	-0.0036	&		&	-0.0026	&		&		&	-0.0032	\\
rs4693369	&	82671234	&	RP11-689K5.3	&	-0.0156	&	-0.0046	&		&	-0.0027	&		&		&	-0.0049	\\
rs12508164	&	82671299	&	RP11-689K5.3	&	-0.012	&	-0.0029	&		&	-0.001	&		&		&	-0.0038	\\
rs7672440	&	82671938	&	RP11-689K5.3	&	-0.0079	&	-0.0013	&		&		&		&		&	-0.0018	\\
rs1353661	&	82672523	&	RP11-689K5.3	&	-0.0025	&		&		&		&		&		&		\\
rs676592	&	82733530	&	RP11-689K5.3	&	-0.0012	&		&		&		&		&		&		\\
rs1993798	&	82762741	&	RP11-689K5.3	&	0.0047	&		&		&		&		&		&		\\
rs2868257	&	82762839	&	RP11-689K5.3	&	0.0072	&		&		&		&		&		&		\\
rs6535281	&	82763010	&	RP11-689K5.3	&	0.0048	&		&		&		&		&		&		\\
rs6535291	&	82926694	&	RP11-689K5.3	&	-0.0011	&		&		&		&		&		&		\\
rs434193	&	86253489	&	ARHGAP24	&	-0.0032	&		&		&		&		&		&		\\
rs6842681	&	86253994	&	ARHGAP24	&	-0.0084	&	0.0019	&		&	-0.0036	&	-0.0033	&		&	-0.0031	\\
rs425196	&	86255297	&	ARHGAP24	&	-0.0144	&	0.0054	&		&	-0.0083	&	-0.0073	&		&	-0.0076	\\
rs416035	&	86255366	&	ARHGAP24	&	-0.0203	&	0.0111	&	-0.0014	&	-0.014	&	-0.0117	&		&	-0.0136	\\
rs432755	&	86255399	&	ARHGAP24	&	-0.0252	&	0.0174	&	-0.003	&	-0.0196	&	-0.016	&		&	-0.0196	\\
rs375432	&	86255845	&	ARHGAP24	&	-0.0276	&	0.0214	&	-0.0041	&	-0.0224	&	-0.0185	&		&	-0.023	\\
rs425642	&	86255997	&	ARHGAP24	&	-0.0257	&	0.0207	&	-0.0037	&	-0.0194	&	-0.016	&		&	-0.021	\\
rs407430	&	86256356	&	ARHGAP24	&	-0.0204	&	0.0151	&	-0.0023	&	-0.0128	&	-0.0109	&		&	-0.0148	\\
rs400023	&	86256538	&	ARHGAP24	&	-0.0131	&	0.0077	&		&	-0.0062	&	-0.0056	&		&	-0.0077	\\
rs585787	&	86257453	&	ARHGAP24	&	-0.006	&	0.0023	&		&	-0.0019	&	-0.0019	&		&	-0.0025	\\
rs380632	&	86264123	&	ARHGAP24	&	-0.0012	&		&		&		&		&		&		\\
rs2726516	&	106346206	&	PPA2	&	0.0022	&		&		&		&		&		&		\\
rs2636739	&	106352105	&	PPA2	&	0.0022	&		&		&		&		&		&		\\
\hline
\end{tabular}
 \begin{tablenotes}
\item $^*$ Genes that SNPs belong to or are the closest to.
 \end{tablenotes}
\end{table}
\end{center}

\begin{center}
\begin{table}
\caption{Analysis of the TCGA SKCM data using the proposed approach: identified main effects and interactions} \label{Tab:SKCM}
\centering
\begin{tabular}{lcccccc}
\hline
~~Gene	&		&	Age	&	PN	&	Gender	&	Breslow's depth	&	Clark level	\\
\hline
	&		&	-0.1381	&	-0.3077	&	0.0536	&	-0.2158	&	-0.1590	\\
ACTL6B	&	-0.0067	&		&		&		&		&		\\
BLOC1S5	&	0.0188	&	-0.0012	&		&		&		&		\\
C3ORF67	&	0.0420	&	-0.0014	&	0.0064	&		&	0.0169	&	0.0016	\\
CLEC2L	&	-0.0069	&		&		&		&		&		\\
CLPB	&	-0.0048	&		&		&		&		&		\\
CREG1	&	0.0281	&		&		&	-0.0018	&		&		\\
CRYBA1	&	-0.0037	&		&		&		&		&		\\
ENDOD1	&	0.0160	&		&	-0.0013	&	-0.0014	&		&	-0.0020	\\
ETV3	&	-0.0019	&		&		&		&		&		\\
FAM131B	&	0.0041	&		&		&		&		&		\\
GOLPH3L	&	-0.0024	&		&		&		&		&		\\
IFNA7	&	-0.0018	&		&		&		&		&		\\
IL17A	&	0.0046	&		&		&		&		&		\\
IL17F	&	0.0143	&		&		&	-0.0014	&		&		\\
IL34	&	0.0030	&		&		&		&		&		\\
INPP5K	&	0.0093	&		&		&		&		&		\\
INTS4	&	-0.0055	&		&		&		&		&		\\
ISL2	&	-0.0022	&		&		&		&		&		\\
KCNE1	&	0.0281	&	-0.0026	&		&	0.0024	&		&	-0.0055	\\
LAMTOR1	&	-0.0078	&		&		&		&		&		\\
LANCL2	&	0.0149	&		&		&		&		&	0.0012	\\
LYNX1	&	-0.0261	&	0.0012	&	0.0021	&	-0.0011	&		&	-0.0026	\\
MEPE	&	0.0144	&	-0.0012	&		&		&		&	-0.0014	\\
METTL21C	&	0.0087	&		&		&		&		&		\\
NKAIN2	&	-0.0239	&	0.0019	&	0.0027	&		&	0.0013	&	0.0033	\\
NKAIN3	&	-0.0019	&		&		&		&		&		\\
NOV	&	0.0422	&		&	-0.0070	&	-0.0085	&	-0.0060	&	0.0047	\\
OR5L2	&	0.0452	&	-0.0103	&	-0.0064	&	-0.0103	&	-0.0037	&	0.0053	\\
PRSS3	&	-0.0100	&		&		&		&		&		\\
PXDNL	&	-0.0106	&		&		&		&		&		\\
RAC1	&	-0.0177	&		&		&		&		&	0.0013	\\
RAET1L	&	0.0093	&		&		&		&		&		\\
RIMS2	&	0.0076	&		&		&		&		&		\\
RPTN	&	-0.0023	&		&		&		&		&		\\
SERPINB13	&	-0.0079	&		&		&		&		&		\\
SERPINB3	&	-0.0018	&		&		&		&		&		\\
\hline \multicolumn{7}{r}{{Continued on the next page}}
\end{tabular}
\end{table}
\end{center}

\setcounter{table}{3}

\begin{center}
\begin{table}
\caption{Continued from the previous page} \label{Tab:SKCM}
\centering
\begin{tabular}{lcccccc}
\hline
~~Gene	&		&	Age	&	PN	&	Gender	&	Breslow's depth	&	Clark level	\\
\hline
SETD3	&	0.0139	&		&		&		&		&		\\
SKIDA1	&	-0.0075	&		&		&		&		&		\\
SLFN13	&	0.0212	&	-0.0023	&		&	-0.0023	&		&		\\
SPINK4	&	0.0012	&		&		&		&		&		\\
SPRR2B	&	0.0015	&		&		&		&		&		\\
STMN4	&	0.0035	&		&		&		&		&		\\
STPG4	&	-0.0055	&		&		&		&		&		\\
SYT12	&	0.0027	&		&		&		&		&		\\
TAS2R1	&	-0.0056	&		&		&		&		&		\\
TRIM46	&	-0.0117	&		&		&		&		&		\\
UBE2V1	&	-0.0317	&	-0.0018	&	0.0045	&		&		&	0.0019	\\
UGT1A7	&	-0.0056	&		&		&		&		&		\\
WDPCP	&	-0.0843	&	-0.0194	&	-0.0322	&	-0.0018	&	0.0206	&	-0.0192	\\
WDR77	&	-0.0137	&		&		&		&		&		\\
\hline
\end{tabular}
\end{table}
\end{center}

\clearpage

\setcounter{table}{0}
\renewcommand\thetable{{A\arabic{table}}}
\setcounter{section}{0}
\renewcommand\thesection{{A\arabic{section}}}

\setcounter{equation}{0}
\renewcommand\theequation{{A\arabic{equation}}}

\setcounter{figure}{0}
\renewcommand\thefigure{{A\arabic{figure}}}

\section*{Appendix}

\subsection*{Estimation under the AFT model}
For subject $i$, denote $T_i$ as the survival time of interest. Use notations similar to those in the main text. For $T_i$, consider the accelerated failure time (AFT) model
\begin{equation*}\label{AFT}
\log(T_i)=\alpha_0+\sum_{k=1}^q Z_{ik}\alpha_k+\sum_{j=1}^p X_{ij} \beta_j+\sum_{k=1}^q\sum_{j=1}^p Z_{ik} X_{ij}\eta_{kj} +\varepsilon_i,
\end{equation*}
where $\alpha_0$ is the intercept. In practice, right censoring is usually present. Denote $C_i$ as the censoring time for subject $i$, then we observe $Y_i=\log(\min(T_i, C_i))$ and $\tilde{\delta}_i=I(T_i\leq C_i)$. Assume that data $\{(\bm{Z}_{i\cdot}, \bm{X}_{i\cdot}, Y_i,\tilde{\delta}_i), i=1, \ldots, n\}$ have been sorted according to $Y_i$ from the smallest to the largest. For estimation, the following weighted least squared loss function is adopted,
\begin{eqnarray}\label{weighted_ls}
\frac{1}{2n}\sum_{i=1}^n w_i \left[Y_i-\left(\alpha_0+\sum_{k=1}^q Z_{ik}\alpha_k+\sum_{j=1}^p X_{ij} \beta_j+\sum_{k=1}^q\sum_{j=1}^p Z_{ik} X_{ij}\eta_{kj}\right)\right]^2,
\end{eqnarray}
where $w_i$'s are the Kaplan-Meier weights defined as
\begin{equation*}\label{censoring_weight}
w_1=\frac{\tilde{\delta}_1}{n}, ~w_i=\frac{\tilde{\delta}_i}{n-i+1}\prod_{l=1}^{i-1}
\Big(\frac{n-l}{n-l+1}\Big)^{\tilde{\delta}_l},i=2,\cdots,n.
\end{equation*}
We center $Y_i$, $\bm{Z}_{i\cdot}$, $\bm{X}_{i\cdot}$, and $\bm{W}_{i\cdot}^{(k)}=(Z_{ik} X_{i1},\cdots,Z_{ik} X_{ip})$ using their weighted means. Specifically,
\begin{equation*}
Y_i=\sqrt{w_{i}}(Y_{i}-\overline{Y}), ~\bm{Z}_{i\cdot}=\sqrt{w_{i}}(\bm{Z}_{i\cdot}-\overline{\bm{Z}}), ~\bm{X}_{i\cdot}=\sqrt{w_{i}}(\bm{X}_{i\cdot}-\overline{\bm{X}}),~\bm{W}_{i\cdot}^{(k)}=\sqrt{w_{i}}\left(\bm{W}_{i\cdot}^{(k)}-\overline{\bm{W}}^{(k)}\right),
\end{equation*}
where $\overline{Y}=\sum_{i=1}^{n}w_{i}Y_{i}/\sum_{i=1}^n w_{i}$, $\overline{\bm{Z}}=\sum_{i=1}^{n}w_{i}\bm{Z}_{i\cdot}/\sum_{i=1}^n w_{i}$, $\overline{\bm{X}}=\sum_{i=1}^{n}w_{i}\bm{X}_{i\cdot}/\sum_{i=1}^n w_{i}$, and $\overline{\bm{W}}^{(k)}=\sum_{i=1}^{n}w_{i}\bm{W}_{i\cdot}^{(k)}/\sum_{i=1}^n w_{i}$.
Then, loss function (\ref{weighted_ls}) can be rewritten as
\begin{equation*}\label{loss_function4}
\frac{1}{2n}\left\| \bm{Y}-\bm{Z}\bm{\alpha}-\bm{X}\bm{\beta}-\sum_{k=1}^q \bm{W}^{(k)}\bm{\eta}_k\right\|_2^2.
\end{equation*}

\clearpage

\subsection*{Details for Steps 2.1 and 2.2 of the proposed algorithm}

Consider the objective function
\begin{eqnarray*}\label{beta2}
\bm{\beta}^{(t)}=\textrm{argmin} \frac{1}{2n}\left\|\tilde{\bm{Y}}^{(t)}-\tilde{\bm{X}}^{(t)}\bm{\beta}\right\|_2^2+\sum_{j=1}^p \rho(|\beta_j|;\lambda_1,r)+\frac{1}{2}\lambda_2\left(\bm{\beta}'\bm{J}\bm{\beta}\right).
\end{eqnarray*}

For $j=1, \ldots, p$, the CD algorithm optimizes the objective function with respect to $\beta_j$ while fixing the other parameters $\beta_l (l\neq j)$ at their current estimates $\beta^{(t)}_l$ for $l<j$ or $\beta^{(t-1)}_l$ for $l>j$. Specifically, consider the following simplified objective function
\begin{equation}\label{loss_b}
Q_s(\beta_j)=\frac{1}{2n}\left\|\bm{res}_{-j}^{(t)}-\tilde{\bm{X}}^{(t)}_j\beta_j\right\|_2^2+\rho(|\beta_j|;\lambda_1,r)+\frac{1}{2}
\lambda_2\left(J_{jj} \beta_j^2+2\sum_{l=1}^{j-1} \beta^{(t)}_l J_{jl} +2\sum_{l=j+1}^p \beta^{(t-1)}_l J_{jl}\right),
\end{equation}
where $\bm{res}_{-j}^{(t)}=\tilde{\bm{Y}}^{(t)}-\sum_{l=1}^{j-1}\tilde{\bm{X}}_{l}^{(t)}\beta_l^{(t)}-\sum_{l=j+1}^p\tilde{\bm{X}}_{l}^{(t)}\beta_l^{(t-1)}$. The first order derivative of (\ref{loss_b}) is
\begin{small}
\begin{eqnarray*}\label{firstder}
\nonumber\frac{\partial Q_s(\beta_j)}{\partial \beta_j}&=&-\frac{1}{n}\left(\bm{\tilde{X}}_{j}^{(t)}\right)'\bm{res}_{-j}^{(t)}+ \frac{1}{n} \left(\bm{\tilde{X}}_{j}^{(t)}\right)'\bm{\tilde{X}}_{j}^{(t)}\beta_j+\lambda_1 sgn(\beta_j)\left\{\begin{array}{ll}
1-\frac{|\beta_j|}{\lambda_1 r} & |\beta_j|\leq \lambda_1 r\\
0 & |\beta_j|> \lambda_1 r,
\end{array}\right.+\lambda_2J_{jj}\beta_j+\lambda_2\Delta_j^{(t)},\\
&\triangleq& -\varphi_j^{(t)}+\chi_j^{(t)}\beta_j+\lambda_1 sgn(\beta_j)\left\{\begin{array}{ll}
1-\frac{|\beta_j|}{\lambda_1 r} & |\beta_j|\leq \lambda_1 r\\
0 & |\beta_j|> \lambda_1 r,
\end{array}\right.+\lambda_2J_{jj}\beta_j+\lambda_2\Delta_j^{(t)},
\end{eqnarray*}
\end{small}
where
\[\varphi_j^{(t)}=\frac{1}{n}\left(\bm{\tilde{X}}_{j}^{(t)}\right)'\bm{res}_{-j}^{(t)},~\chi_j^{(t)}=\frac{1}{n} \left(\bm{\tilde{X}}_{j}^{(t)}\right)'\bm{\tilde{X}}_{j}^{(t)},~\Delta_j^{(t)}=\sum_{l=1}^{j-1} \beta^{(t)}_l J_{jl}+\sum_{l=j+1}^p \beta^{(t-1)}_l J_{jl}.
\]
By setting the first order derivative equal to zero, we have
\[\beta_j^{(t)}=\left\{\begin{array}{ll}
\frac{\textrm{ST}\left(\varphi_j^{(t)}-\lambda_2\Delta_j^{(t)},\lambda_1\right)}{\chi_j^{(t)}+\lambda_2J_{jj}-\frac{1}{r}} & \left|\varphi_j^{(t)}-\lambda_2\Delta_j^{(t)}\right|\leq\lambda_1r(\chi_j^{(t)}+\lambda_2J_{jj})\\
\frac{\varphi_j^{(t)}-\lambda_2\Delta_j}{\chi_j^{(t)}+\lambda_2J_{jj}} & \left|\varphi_j^{(t)}-\lambda_2\Delta_j^{(t)}\right|>\lambda_1r(\chi_j^{(t)}+\lambda_2J_{jj})
\end{array}\right.,\]
where $\textrm{ST}(\nu,\lambda_1)=sgn(\nu)(|\nu|-\lambda_1)_+$ is the soft-thresholding operator.

\clearpage

\subsection*{Proof of Theorem 1}

To prove Theorem 1, it suffices to show that under conditions (C1)-(C5), for a given $\xi$,
\[P\left\{\underset{\bm{\theta}_{\mathcal{A}}\in \mathcal{N}_1}{\inf} \tilde{Q}_n(\bm{\theta}_{\mathcal{A}})>\tilde{Q}_n(\bm{\theta}^0_{\mathcal{A}})\right\}\geq 1-\xi,\]
where $\mathcal{N}_1=\{\bm{\theta}_{\mathcal{A}}:||\bm{\theta}_{\mathcal{A}}-\bm{\theta}^0_{\mathcal{A}}||_2=\delta_n\}$.

Let $\bm{w}=\left(\bm{g}_{q\times1}',\bm{u}_{|\mathcal{A}_1|\times1}',\bm{v}_{1_{|\mathcal{A}_2^1|\times1}}',\cdots,\bm{v}_{q_{|\mathcal{A}_2^q|\times1}}'\right)'$ with $||\bm{w}||_2=1$ and $\bm{\theta}_{\mathcal{A}}=\bm{\theta}_{\mathcal{A}}^0+\delta_n \bm{w}$.
Let $L_n(\bm{\theta}_{\mathcal{A}})=\left\| \bm{Y}-\bm{Z}\bm{\alpha}-\bm{X}_{\mathcal{A}_1}\bm{\beta}_{\mathcal{A}_1}-\sum_{k=1}^q \bm{W}^{(k)}_{\mathcal{A}_2^k}(\bm{\beta}_{\mathcal{A}_2^k}\odot \bm{\gamma}_{k,\mathcal{A}_2^k})\right\|_2^2$, then
\begin{eqnarray*}
\nonumber D_n(\bm{w})&=&\tilde{Q}_n(\bm{\theta}_{\mathcal{A}}^0+\delta_n \bm{w})-\tilde{Q}_n(\bm{\theta}_{\mathcal{A}}^0)\\
\nonumber &=& \frac{1}{2n}L_n(\bm{\theta}_{\mathcal{A}}^0+\delta_n \bm{w})-\frac{1}{2n}L_n(\bm{\theta}_{\mathcal{A}}^0)\\
\nonumber &+&\frac{1}{2}\lambda_2(\bm{\beta}_{\mathcal{A}_1}^0+\delta_n \bm{u})'\bm{J}_{\mathcal{A}_1,\mathcal{A}_1}(\bm{\beta}_{\mathcal{A}_1}^0+\delta_n \bm{u})-\frac{1}{2}\lambda_2(\bm{\beta}_{\mathcal{A}_1}^0)'\bm{J}_{\mathcal{A}_1,\mathcal{A}_1}\bm{\beta}_{\mathcal{A}_1}^0\\
 &+&\frac{1}{2}\lambda_2\sum_{k=1}^q(\bm{\gamma}^0_{\mathcal{A}_2^k}+\delta_n \bm{v}_k)'\bm{J}_{\mathcal{A}_2^k,\mathcal{A}_2^k}(\bm{\gamma}^0_{\mathcal{A}_2^k}+\delta_n \bm{v}_k)-\frac{1}{2}\lambda_2\sum_{k=1}^q(\bm{\gamma}^0_{\mathcal{A}_2^k})'\bm{J}_{\mathcal{A}_2^k,\mathcal{A}_2^k}\bm{\gamma}_{\mathcal{A}_2^k}^0.
\end{eqnarray*}
We have
\begin{eqnarray*}
I&\triangleq&\frac{1}{2n}L_n(\bm{\theta}_{\mathcal{A}}^0+\delta_n \bm{w})-\frac{1}{2n}L_n(\bm{\theta}_{\mathcal{A}}^0)\\
&=&\frac{1}{2n}\delta_n \bm{w}'(\left.\nabla L_n(\bm{\theta}_{\mathcal{A}})\right|_{\bm{\theta}_{\mathcal{A}}^0})+\frac{1}{4n}\delta_n^2\bm{w}'(\left.\nabla^2 L_n(\bm{\theta}_{\mathcal{A}})\right|_{\tilde{\bm{\theta}}_{\mathcal{A}}})\bm{w}\\
&=&\delta_n\bm{w}'\left[-\frac{1}{n}\bm{G}(\bm{\beta}^0_{\mathcal{A}_2},\bm{\gamma}^0_{\mathcal{A}_1})'\bm{\varepsilon}\right]\\
&+&\frac{1}{2}\delta_n^2 \bm{w}'\left[\frac{1}{n}\bm{G}(\tilde{\bm{\beta}}_{\mathcal{A}_2},\tilde{\bm{\gamma}}_{\mathcal{A}_1})'\bm{G}(\tilde{\bm{\beta}}_{\mathcal{A}_2},\tilde{\bm{\gamma}}_{\mathcal{A}_1})+\frac{1}{n}\bm{F}(\tilde{\bm{\theta}}_{\mathcal{A}})\right]\bm{w}
\\
&\triangleq&I_1+I_2,
\end{eqnarray*}
where $\bm{\varepsilon}=\bm{Y}-\bm{Z}\bm{\alpha}^0-\bm{X}_{\mathcal{A}_1}\bm{\beta}^0_{\mathcal{A}_1}-\sum_{k=1}^q \bm{W}^{(k)}_{\mathcal{A}_2^k}(\bm{\beta}^0_{\mathcal{A}_2^k}\odot \bm{\gamma}^0_{k,\mathcal{A}_2^k})$,  $\bm{\gamma}_{\mathcal{A}_1}=\left(\bm{\gamma}'_{1,\mathcal{A}_1},\cdots,\bm{\gamma}'_{q,\mathcal{A}_1}\right)'$ with $\gamma_{kj}=0$, if $j\in \mathcal{A}_1$ but $j\notin \mathcal{A}_2^k$,
\[\bm{G}(\bm{\beta}_{\mathcal{A}_2},\bm{\gamma}_{\mathcal{A}_1})=\left(\bm{Z}, \bm{U}(\bm{\gamma}_{\mathcal{A}_1}), \bm{V}^{(1)}(\bm{\beta}_{\mathcal{A}_2^1}),\bm{V}^{(2)}(\bm{\beta}_{\mathcal{A}_2^2}),\cdots, \bm{V}^{(q)}(\bm{\beta}_{\mathcal{A}_2^q})\right)_{n\times (q+s)},
\]
with \[\bm{U}(\bm{\gamma}_{\mathcal{A}_1})=\bm{X}_{\mathcal{A}_1}+\sum_{k=1}^q \bm{W}_{\mathcal{A}_1}^{(k)}\odot \left(\bm{1}_{n\times 1}\left(\bm{\gamma}_{k,{\mathcal{A}_1}}\right)'\right),~\bm{V}^{(k)}(\bm{\beta}_{\mathcal{A}_2^k})=\bm{W}^{(k)}_{\mathcal{A}_2^k}\odot \left(\bm{1}_{n\times 1}\left(\bm{\beta}_{\mathcal{A}_2^k}\right)'\right),\]
$\bm{F}(\bm{\theta}_{\mathcal{A}})=(f_{jl}(\bm{\theta}_{\mathcal{A}}))_{(q+s)\times(q+s)}$ with $f_{jl}(\bm{\theta}_{\mathcal{A}})=-\left(\bm{W}_{\varsigma}^{(k)}\right)'(\bm{Y}-\bm{Z}\bm{\alpha}-\bm{X}_{\mathcal{A}_1}\bm{\beta}_{\mathcal{A}_1}-\sum_{g=1}^q \bm{W}^{(g)}_{\mathcal{A}_2^{g}}(\bm{\beta}_{\mathcal{A}_2^{g}}\odot \bm{\gamma}_{g,\mathcal{A}_2^{g}}))$ if both $j$ and $l$ correspond to the $\varsigma$th element of $\mathcal{A}_2^k$, and 0 otherwise, and $\tilde{\bm{\theta}}_{\mathcal{A}}$ lies on the line segment joining $\bm{\theta}_{\mathcal{A}}$ and $\bm{\theta}^0_{\mathcal{A}}$. Moreover,
\begin{eqnarray*}
\nonumber II&\triangleq&\frac{1}{2}\lambda_2(\bm{\beta}_{\mathcal{A}_1}^0+\delta_n \bm{u})'\bm{J}_{\mathcal{A}_1,\mathcal{A}_1}(\bm{\beta}_{\mathcal{A}_1}^0+\delta_n \bm{u})-\frac{1}{2}\lambda_2(\bm{\beta}_{\mathcal{A}_1}^0)'\bm{J}_{\mathcal{A}_1,\mathcal{A}_1}\bm{\beta}_{\mathcal{A}_1}^0\\
\nonumber &+&\frac{1}{2}\lambda_2\sum_{k=1}^q(\bm{\gamma}^0_{\mathcal{A}_2^k}+\delta_n \bm{v}_k)'\bm{J}_{\mathcal{A}_2^k,\mathcal{A}_2^k}(\bm{\gamma}^0_{\mathcal{A}_2^k}+\delta_n \bm{v}_k)-\frac{1}{2}\lambda_2\sum_{k=1}^q(\bm{\gamma}^0_{\mathcal{A}_2^k})'\bm{J}_{\mathcal{A}_2^k,\mathcal{A}_2^k}\bm{\gamma}_{\mathcal{A}_2^k}^0\\
\nonumber &=& \delta_n \lambda_2\bm{w}'\widetilde{\bm{J}}_{\mathcal{A},\mathcal{A}}\bm{\theta}^0_{\mathcal{A}}+\frac{1}{2}\delta_n^2 \lambda_2 \bm{w}'\widetilde{\bm{J}}_{\mathcal{A},\mathcal{A}}\bm{w}\\
%&\geq& -\delta_n \bar{\lambda}^{\ast}|\widetilde{\bm{J}}_{\mathcal{A},\mathcal{A}}\bm{\theta}^0_{\mathcal{A}}|_2+II_2
&\geq& -\delta_n \lambda_2||\widetilde{\bm{J}}_{\mathcal{A},\mathcal{A}}\bm{\theta}^0_{\mathcal{A}}||_2,
\end{eqnarray*}
where $\widetilde{\bm{J}}_{\mathcal{A},\mathcal{A}}=\textrm{diag}\left(\bm{0}_{q\times q}, \bm{J}_{\mathcal{A}_1,\mathcal{A}_1}, \cdots, \bm{J}_{\mathcal{A}_2^q,\mathcal{A}_2^q}\right)$ is a block diagonal matrix with the diagonal blocks being $\bm{0}_{q\times q}, \bm{J}_{\mathcal{A}_1,\mathcal{A}_1}, \cdots,$ and $\bm{J}_{\mathcal{A}_2^q,\mathcal{A}_2^q}$, and
$\frac{1}{2}\delta_n^2 \lambda_2\bm{w}'\widetilde{\bm{J}}_{\mathcal{A},\mathcal{A}}\bm{w}\geq \frac{1}{2}\delta_n^2\lambda_2\lambda_{\min}\left(\widetilde{\bm{J}}_{\mathcal{A},\mathcal{A}}\right)\geq0$ with condition (C5).

With $\delta_n=\frac{4\lambda_2||\widetilde{\bm{J}}_{\mathcal{A},\mathcal{A}}\bm{\theta}^0_{\mathcal{A}}||_2}{\underline{c}}+E\sqrt{s/n}$, and conditions (C2), (C4) and (C5), we have
\begin{equation*}
||\tilde{\bm{\theta}}_{\mathcal{A}}-\bm{\theta}_{\mathcal{A}}^0||_{\infty}\leq ||\bm{\theta}_{\mathcal{A}}-\bm{\theta}_{\mathcal{A}}^0||_{\infty}\leq\delta_n<b_0/2.
\end{equation*}
Then, with condition (C3), we have
\begin{equation*}
I_2\geq \frac{1}{2}\delta_n^2\underline{c}>0.
\end{equation*}

For $I_1$, with conditions (C1) and (C3), we have
\begin{eqnarray*}
&&P\left(\delta_n\bm{w}'\left[-\frac{1}{n}\bm{G}(\bm{\beta}^0_{\mathcal{A}_2},\bm{\gamma}^0_{\mathcal{A}_1})'\bm{\varepsilon}\right]\leq -\delta_n \epsilon\right)\\
&=&P\left(\frac{\bm{w}'\left[-\frac{1}{n}\bm{G}(\bm{\beta}^0_{\mathcal{A}_2},\bm{\gamma}^0_{\mathcal{A}_1})'\bm{\varepsilon}\right]}{\left\|\bm{w}'\left[-\frac{1}{n}\bm{G}(\bm{\beta}^0_{\mathcal{A}_2},\bm{\gamma}^0_{\mathcal{A}_1})'\right]\right\|_2}\leq -\frac{\epsilon}{\left\|\bm{w}'\left[-\frac{1}{n}\bm{G}(\bm{\beta}^0_{\mathcal{A}_2},\bm{\gamma}^0_{\mathcal{A}_1})'\right]\right\|_2}\right)\\
&\leq & \exp\left(-\frac{n\epsilon^2}{2\sigma^2\bar{c}s}\right).
\end{eqnarray*}
Set $\epsilon=\frac{1}{4}\underline{c}\delta_n$, we have
\begin{eqnarray*}
P\left(\delta_n\bm{w}'\left[-\frac{1}{n}\bm{G}(\bm{\beta}^0_{\mathcal{A}_2},\bm{\gamma}^0_{\mathcal{A}_1})'\bm{\varepsilon}\right]\geq -\frac{1}{4}\underline{c}\delta_n^2\right)\geq 1- \exp\left(-\frac{n\underline{c}^2\delta_n^2}{32\sigma^2\bar{c}s}\right).
\end{eqnarray*}
Thus, with $\delta_n=\frac{4\lambda_2||\widetilde{\bm{J}}_{\mathcal{A},\mathcal{A}}\bm{\theta}^0_{\mathcal{A}}||_2}{\underline{c}}+E\sqrt{s/n}$, we have
\begin{eqnarray*}
P\left\{\underset{\hat{\theta}\in \mathcal{N}_1}{\inf} Q_n(\hat{\bm{\theta}})>Q_n(\bm{\theta}^0)\right\}&\geq& P\left\{D_n(\bm{w})> 0\right\}\\
&\geq & P\left\{\delta_n\bm{w}'\left[-\frac{1}{n}\bm{G}(\bm{\beta}^0_{\mathcal{A}_2},\bm{\gamma}^0_{\mathcal{A}_1})'\bm{\varepsilon}\right]+\frac{1}{2}\delta_n^2 \underline{c}-\delta_n \lambda_2||\widetilde{\bm{J}}_{\mathcal{A},\mathcal{A}}\bm{\theta}^0_{\mathcal{A}}||_2>0  \right\}\\
&\geq& P\left(\delta_n\bm{w}'\left[-\frac{1}{n}\bm{G}(\bm{\beta}^0_{\mathcal{A}_2},\bm{\gamma}^0_{\mathcal{A}_1})'\bm{\varepsilon}\right]\geq -\frac{1}{4}\underline{c}\delta_n^2\right)\\
&\geq& 1-\exp\left(-\frac{n\underline{c}^2\delta_n^2}{32\sigma^2\bar{c}s}\right)\\
&=&1-\exp\left(-\frac{\left[4\sqrt{n/s}\lambda_2||\widetilde{\bm{J}}_{\mathcal{A},\mathcal{A}}\bm{\theta}^0_{\mathcal{A}}||_2+E\underline{c}\right]^2}{32\sigma^2\bar{c}}\right).
\end{eqnarray*}
This completes the proof of Theorem 1.

\clearpage

\subsection*{Proof of Theorem 2}

First, consider $\hat{\bm{\beta}}_{\mathcal{A}_1^c}$. Following Theorem 1 in \cite{Fan11}, with condition (C9) and Theorem 1, it suffices to check condition (8) in \cite{Fan11}. Let
\[h_1=(n\lambda_1)^{-1}\left[\frac{1}{2}\left.\nabla_{\bm{\beta}_{\mathcal{A}_1^c}} L_n(\bm{\theta})\right|_{\hat{\bm{\theta}}}+\lambda_2n \bm{J}_{\mathcal{A}_1^c\cdot}\hat{\bm{\beta}}\right].\]
Since $\hat{\bm{\beta}}_{\mathcal{A}_1^c}=0$, with a Taylor expansion, we have
\begin{eqnarray*}
h_1&=&(n\lambda_1)^{-1}\left[-\bm{U}(\bm{\gamma}_{\mathcal{A}_1^c})'\left(\bm{Y}-\bm{Z}\hat{\bm{\alpha}}-\bm{X}\hat{\bm{\beta}}-\sum_{k=1}^q \bm{W}^{(k)}(\hat{\bm{\beta}}\odot \hat{\bm{\gamma}_k})\right)
+\lambda_2n \bm{J}_{\mathcal{A}_1^c,\mathcal{A}_1}\hat{\bm{\beta}}_{\mathcal{A}_1}\right]\\
&=& (n\lambda_1)^{-1}\left[-\bm{U}(\bm{\gamma}^0_{\mathcal{A}_1^c})'\bm{\varepsilon}+\bm{U}(\bm{\gamma}^0_{\mathcal{A}_1^c})'
\bm{G}(\bm{\beta}^0_{\mathcal{A}_2},\bm{\gamma}^0_{\mathcal{A}_1})'(\hat{\bm{\theta}}_{\mathcal{A}}-\bm{\theta}_{\mathcal{A}}^0)+
\bm{\kappa}
+\lambda_2n \bm{J}_{\mathcal{A}_1^c,\mathcal{A}_1}\hat{\bm{\beta}}_{\mathcal{A}_1}\right]\\
&=& (n\lambda_1)^{-1}\left[-\bm{U}(\bm{\gamma}^0_{\mathcal{A}_1^c})'\bm{\varepsilon}+III+\lambda_2n \bm{J}_{\mathcal{A}_1^c,\mathcal{A}_1}\hat{\bm{\beta}}_{\mathcal{A}_1}\right].
\end{eqnarray*}
For $III$, let
$m_j(\bm{\theta}_{\mathcal{A}})=\left(\bm{X}_j+\sum_{k=1}^q\bm{W}_j^{(k)} \gamma_{kj}\right)'\left(\bm{Z}\bm{\alpha}+\bm{X}_{\mathcal{A}_1}\bm{\beta}_{\mathcal{A}_1}+\sum_{k=1}^q \bm{W}^{(k)}_{\mathcal{A}_2^k}(\bm{\beta}_{\mathcal{A}_2^k}\odot \bm{\gamma}_{k,\mathcal{A}_2^k})\right)$. Then $\bm{\kappa}=(\kappa_j,j\in \mathcal{A}_1^c)'$ with
\begin{eqnarray*}
\kappa_j&=&\frac{1}{2}(\hat{\bm{\theta}}_{\mathcal{A}}-\bm{\theta}_{\mathcal{A}}^0)\left(\left.\nabla^2_{\theta_{\mathcal{A}}}m_j(\bm{\theta}_{\mathcal{A}})\right|_{\tilde{\bm{\theta}}_{\mathcal{A}}}\right)(\hat{\bm{\theta}}_{\mathcal{A}}-\bm{\theta}_{\mathcal{A}}^0),\\
&\leq & \max_{j}\frac{1}{2}\lambda_{\max}\left(\bm{T}_1^{(j)}(\tilde{\bm{\gamma}}_{j})\right)||\bm{\theta}^{\ast}_{\mathcal{A}}-\bm{\theta}^0_{\mathcal{A}}||_2,
\end{eqnarray*}
where $\tilde{\bm{\theta}}_{\mathcal{A}}$ lies on the line segment jointing $\theta^{\ast}_{\mathcal{A}}$ and $\theta^0_{\mathcal{A}}$. Here $\bm{T}_1^{(j)}(\bm{\gamma}_{j})=\left(t_{lh}^{(j)}(\bm{\gamma}_{j})\right)_{(q+s)\times(q+s)}$ with $t_{lh}^{(j)}(\bm{\gamma}_{j})=\left(\bm{X}_j+\sum_{g=1}^q\bm{W}_j^{(g)}\gamma_{gj}\right)' \bm{W}_{\varsigma}^{(k)}$, if both $l$ and $h$ correspond to the $\varsigma$th element of $\mathcal{A}_2^k$, and 0 otherwise.
Consider the event
\[\Omega_1=\left\{||\bm{U}(\bm{\gamma}^0_{\mathcal{A}_1^c})'\bm{\varepsilon}||_{\infty}\leq \zeta_n \sqrt{n}\right\},\]
with $\zeta_n=n^a(\log(n))^{1/2}$. With conditions (C6) and (C7), we have
\begin{eqnarray*}
P(\Omega_1)&=&1-P\left\{||\bm{U}(\bm{\gamma}^0_{\mathcal{A}_1^c})'\bm{\varepsilon}||_{\infty}> \zeta_n \sqrt{n}\right\}\\
&\geq& 1-\sum_{j\in \mathcal{A}_1^c} P\left\{||\bm{U}(\bm{\gamma}^0_{j})'\bm{\varepsilon}||> \zeta_n \sqrt{n}\right\}\\
&\geq & 1-2(p-s_0)\exp\left(-\frac{\zeta_n^2 n}{2\sigma^2\max_{j\in \mathcal{A}_1^c}||\bm{U}(\bm{\gamma}^0_{j})||_2^2}\right)\\
&\geq & 1-2p\exp\left(-\frac{\zeta_n^2 n}{2\sigma^2\max_{j\in \mathcal{A}_1^c}||\bm{U}(\bm{\gamma}^0_{j})||_2^2}\right)\rightarrow 1,
\end{eqnarray*}
as $\log(p)=O(n^a)$ and $||\bm{U}(\bm{\gamma}^0_{j})||_2=O(\sqrt{n})$. Thus, with probability tending to 1,
\[||\bm{U}(\bm{\gamma}^0_{\mathcal{A}_1^c})'\bm{\varepsilon}||_{\infty}=O(n^{a/2+1/2}\sqrt{\log n}).\]
Then, condition (C8) gives
\[(n\lambda_1)^{-1}||\bm{U}(\bm{\gamma}^0_{\mathcal{A}_1^c})'\bm{\varepsilon}||_{\infty}=o(1).\]
For $III$, with conditions (C6) and (C8),
\begin{eqnarray*}
(n\lambda_1)^{-1}||III||_{\infty}&=&(n\lambda_1)^{-1}\left[||\bm{U}(\bm{\gamma}^0_{\mathcal{A}_1^c})'\bm{G}(\bm{\beta}^0_{\mathcal{A}_2},
\bm{\gamma}^0_{\mathcal{A}_1})'(\hat{\bm{\theta}}_{\mathcal{A}}-\bm{\theta}_{\mathcal{A}}^0)||_{\infty}
+||\bm{\kappa}||_{\infty}\right]\\
&=& (n\lambda_1)^{-1}\left[O(n)||\bm{\theta}^{\ast}_{\mathcal{A}}-\bm{\theta}^0_{\mathcal{A}}||_2+O(n)||\bm{\theta}^{\ast}_{\mathcal{A}}-\bm{\theta}^0_{\mathcal{A}}||_2^2\right]\\
&=& O(\lambda_1^{-1}\sqrt{s/n})=o(1).
\end{eqnarray*}
With conditions (C4), (C5) and (C8),
\begin{eqnarray*}
(n\lambda_1)^{-1}||\lambda_2n \bm{J}_{\mathcal{A}_1^c,\mathcal{A}_1}\hat{\bm{\beta}}_{\mathcal{A}_1}||_{\infty}&=&(\lambda_1)^{-1}||\lambda_2 \bm{J}_{\mathcal{A}_1^c,\mathcal{A}_1}\bm{\beta}^0_{\mathcal{A}_1}-\lambda_2 \bm{J}_{\mathcal{A}_1^c,\mathcal{A}_1}(\hat{\bm{\beta}}_{\mathcal{A}_1}-\bm{\beta}^0_{\mathcal{A}_1})||_{\infty}\\
&\leq &(\lambda_1)^{-1}||\lambda_2 \bm{J}_{\mathcal{A}_1^c,\mathcal{A}_1}\bm{\beta}^0_{\mathcal{A}_1}||_{\infty}+(\lambda_1)^{-1}||\lambda_2 \bm{J}_{\mathcal{A}_1^c,\mathcal{A}_1}(\hat{\bm{\beta}}_{\mathcal{A}_1}-\bm{\beta}^0_{\mathcal{A}_1})||_{\infty}\\
&=&O(\lambda_1^{-1}\sqrt{s/n})=o(1).
\end{eqnarray*}

Next, consider $\hat{\bm{\gamma}}_{k,(\tilde{\mathcal{A}}_2^k)^c}$. A similar process is adopted  to check condition (8) in \cite{Fan11}.
Let
\[h_2=(n\lambda_1)^{-1}\left[\left.\frac{1}{2}\nabla_{(\tilde{\mathcal{A}}_2^k)^c} L_n(\bm{\theta})\right|_{\hat{\bm{\theta}}}+\lambda_2n \bm{J}_{(\tilde{\mathcal{A}}_2^k)^c\cdot}\hat{\bm{\gamma}}_k\right].\]
Since $\hat{\bm{\gamma}}_{(\tilde{\mathcal{A}}_2^k)^c}=0$ and $\hat{\bm{\beta}}_{(\tilde{\mathcal{A}}_2^k)^c}\neq0$, with a Taylor expansion, we have
\begin{eqnarray*}
h_2&=&(n\lambda_1)^{-1}\left[-\bm{V}^{(k)}(\bm{\beta}_{(\tilde{\mathcal{A}}_2^k)^c})'\left(\bm{Y}-\bm{Z}\hat{\bm{\alpha}}-\bm{X}\hat{\bm{\beta}}-\sum_{k=1}^q \bm{W}^{(k)}(\hat{\bm{\beta}}\odot \hat{\bm{\gamma}_k})\right)
+\lambda_2n \bm{J}_{(\tilde{\mathcal{A}}_2^k)^c\cdot}\hat{\bm{\gamma}}_k \right]\\
&=& (n\lambda_1)^{-1}\left[-\bm{V}^{(k)}(\bm{\beta}^0_{(\tilde{\mathcal{A}}_2^k)^c})'\bm{\varepsilon}+\bm{V}^{(k)}(\bm{\beta}^0_{(\tilde{\mathcal{A}}_2^k)^c})'
\bm{G}(\bm{\beta}^0_{\mathcal{A}_2},\bm{\gamma}^0_{\mathcal{A}_1})'(\hat{\bm{\theta}}_{\mathcal{A}}-\bm{\theta}_{\mathcal{A}}^0)+
\tilde{\bm{\kappa}}
+\lambda_2n \bm{J}_{(\tilde{\mathcal{A}}_2^k)^c,\cdot}\hat{\bm{\gamma}}_k\right]\\
&=& (n\lambda_1)^{-1}\left[-\bm{V}^{(k)}(\beta^0_{(\tilde{\mathcal{A}}_2^k)^c})'\bm{\varepsilon}+IV+\lambda_2n \bm{J}_{(\tilde{\mathcal{A}}_2^k)^c,\mathcal{A}_2^k}\hat{\bm{\gamma}}_{k,\mathcal{A}_2^k}\right].
\end{eqnarray*}
For $IV$, let
$\tilde{m}_j(\bm{\theta}_{\mathcal{A}})=\left(\bm{W}_j^{(k)}\beta_j\right)'\left(\bm{Z}\bm{\alpha}+\bm{X}_{\mathcal{A}_1}\bm{\beta}_{\mathcal{A}_1}+\sum_{k=1}^q \bm{W}^{(k)}_{\mathcal{A}_2^k}(\bm{\beta}_{\mathcal{A}_2^k}\odot \bm{\gamma}_{k,\mathcal{A}_2^k})\right)$, then $\tilde{\bm{\kappa}}=(\tilde{\kappa}_j,j\in (\tilde{\mathcal{A}}_2^k)^c)'$ with
\begin{eqnarray*}
\tilde{\kappa}_j&=&\frac{1}{2}(\hat{\bm{\theta}}_{\mathcal{A}}-\bm{\theta}_{\mathcal{A}}^0)\left(\left.\nabla^2_{\theta_{\mathcal{A}}}\tilde{m}_j(\bm{\theta}_{\mathcal{A}})\right|_{\tilde{\bm{\theta}}_{\mathcal{A}}}\right)(\hat{\bm{\theta}}_{\mathcal{A}}-\bm{\theta}_{\mathcal{A}}^0),\\
&\leq & \max_{j}\frac{1}{2}\lambda_{\max}\left(\bm{T}_2^{(j)}(\tilde{\beta}_j)\right)||\bm{\theta}^{\ast}_{\mathcal{A}}-\bm{\theta}^0_{\mathcal{A}}||_2,
\end{eqnarray*}
where $\tilde{\bm{\theta}}_{\mathcal{A}}$ lies on the line segment jointing $\bm{\theta}^{\ast}_{\mathcal{A}}$ and $\bm{\theta}^0_{\mathcal{A}}$.
Here $\bm{T}_2^{(j)}(\beta_j)=\left(t_{lh}^{(j)}(\beta_j)\right)_{(q+s)\times (q+s)}$ with $t_{lh}^{(j)}(\beta_j)=\left(\bm{W}_j^{(k)}\beta_j\right)' \bm{W}_{\varsigma}^{(k)}$ if both $l$ and $h$ correspond to the $\varsigma$th element of $\mathcal{A}_2^k$, and 0 otherwise.

Consider the event
\[\Omega_2=\left\{||\bm{V}^{(k)}(\bm{\beta}^0_{(\tilde{\mathcal{A}}_2^k)^c})'\bm{\varepsilon}||_{\infty}\leq \zeta_n \sqrt{n}\right\},\]
with $\zeta_n=n^a(\log(n))^{1/2}$. We have
\begin{eqnarray*}
P(\Omega_2)&=&1-P\left\{||\bm{V}^{(k)}(\bm{\beta}^0_{(\tilde{\mathcal{A}}_2^k)^c})'\bm{\varepsilon}||_{\infty}> \zeta_n \sqrt{n}\right\}\\
&\geq& 1-\sum_{j\in (\tilde{\mathcal{A}}_2^k)^c} P\left\{||\bm{V}^{(k)}(\beta^0_j)'\bm{\varepsilon}||> \zeta_n \sqrt{n}\right\}\\
&\geq & 1-2p\exp\left(-\frac{\zeta_n^2 n}{2\sigma^2\max_{j\in (\tilde{\mathcal{A}}_2^k)^c}||\bm{V}^{(k)}(\beta^0_j)||_2^2}\right)\rightarrow 1,\\
\end{eqnarray*}
as $\log(p)=O(n^a)$ and $||\bm{V}^{(k)}(\beta^0_j)||_2=O(\sqrt{n})$. Thus, we have, with probability tending to 1,
\[||\bm{V}^{(k)}(\bm{\beta}^0_{(\tilde{\mathcal{A}}_2^k)^c})'\bm{\varepsilon}||_{\infty}=O(n^{a/2+1/2}\sqrt{\log n}).\]
The condition (C8) gives
\[(n\lambda_1)^{-1}||\bm{V}^{(k)}(\bm{\beta}^0_{(\tilde{\mathcal{A}}_2^k)^c})'\bm{\varepsilon}||_{\infty}=o(1).\]
For $IV$, with conditions (C6) and (C8),
\begin{eqnarray*}
(n\lambda_1)^{-1}||IV||_{\infty}&=&(n\lambda_1)^{-1}\left[||\bm{V}^{(k)}(\bm{\beta}^0_{(\tilde{\mathcal{A}}_2^k)^c})'
\bm{G}(\bm{\beta}^0_{\mathcal{A}_2},\bm{\gamma}^0_{\mathcal{A}_1})'(\hat{\bm{\theta}}_{\mathcal{A}}-\bm{\theta}_{\mathcal{A}}^0)||_{\infty}
+||\tilde{\bm{\kappa}}||_{\infty}\right]\\
&=& (n\lambda_1)^{-1}\left[O(n)||\bm{\theta}^{\ast}_{\mathcal{A}}-\bm{\theta}^0_{\mathcal{A}}||_2+O(n)||\bm{\theta}^{\ast}_{\mathcal{A}}-\bm{\theta}^0_{\mathcal{A}}||_2^2\right]\\
&=& O(\lambda_1^{-1}\sqrt{s/n})=o(1).
\end{eqnarray*}
With conditions (C4), (C5) and (C8),
\begin{eqnarray*}
(\lambda_1)^{-1}||\lambda_2 \bm{J}_{(\tilde{\mathcal{A}}_2^k)^c,\mathcal{A}_2^k}\hat{\bm{\gamma}}_{k,\mathcal{A}_2^k}||_{\infty}&=&(\lambda_1)^{-1}||\lambda_2 \bm{J}_{(\tilde{\mathcal{A}}_2^k)^c,\mathcal{A}_2^k}\bm{\gamma}^0_{k,\mathcal{A}_2^k}-\lambda_2 \bm{J}_{(\tilde{\mathcal{A}}_2^k)^c,\mathcal{A}_2^k}(\hat{\bm{\gamma}}_{k,\mathcal{A}_2^k}-\bm{\gamma}^0_{k,\mathcal{A}_2^k})||_{\infty}\\
&\leq &(\lambda_1)^{-1}||\lambda_2 \bm{J}_{(\tilde{\mathcal{A}}_2^k)^c,\mathcal{A}_2^k}\bm{\gamma}^0_{k,\mathcal{A}_2^k}||_{\infty}+(\lambda_1)^{-1}||\lambda_2 \bm{J}_{(\tilde{\mathcal{A}}_2^k)^c,\mathcal{A}_2^k}(\hat{\bm{\gamma}}_{k,\mathcal{A}_2^k}-\bm{\gamma}^0_{k,\mathcal{A}_2^k})||_{\infty}\\
&=&O(\lambda_1^{-1}\sqrt{s/n})=o(1).
\end{eqnarray*}
This completes the proof.

\clearpage

\subsection*{Additional numerical results}

\begin{figure}[H]
  \centering
  \includegraphics[width=0.7\textwidth, angle=0]{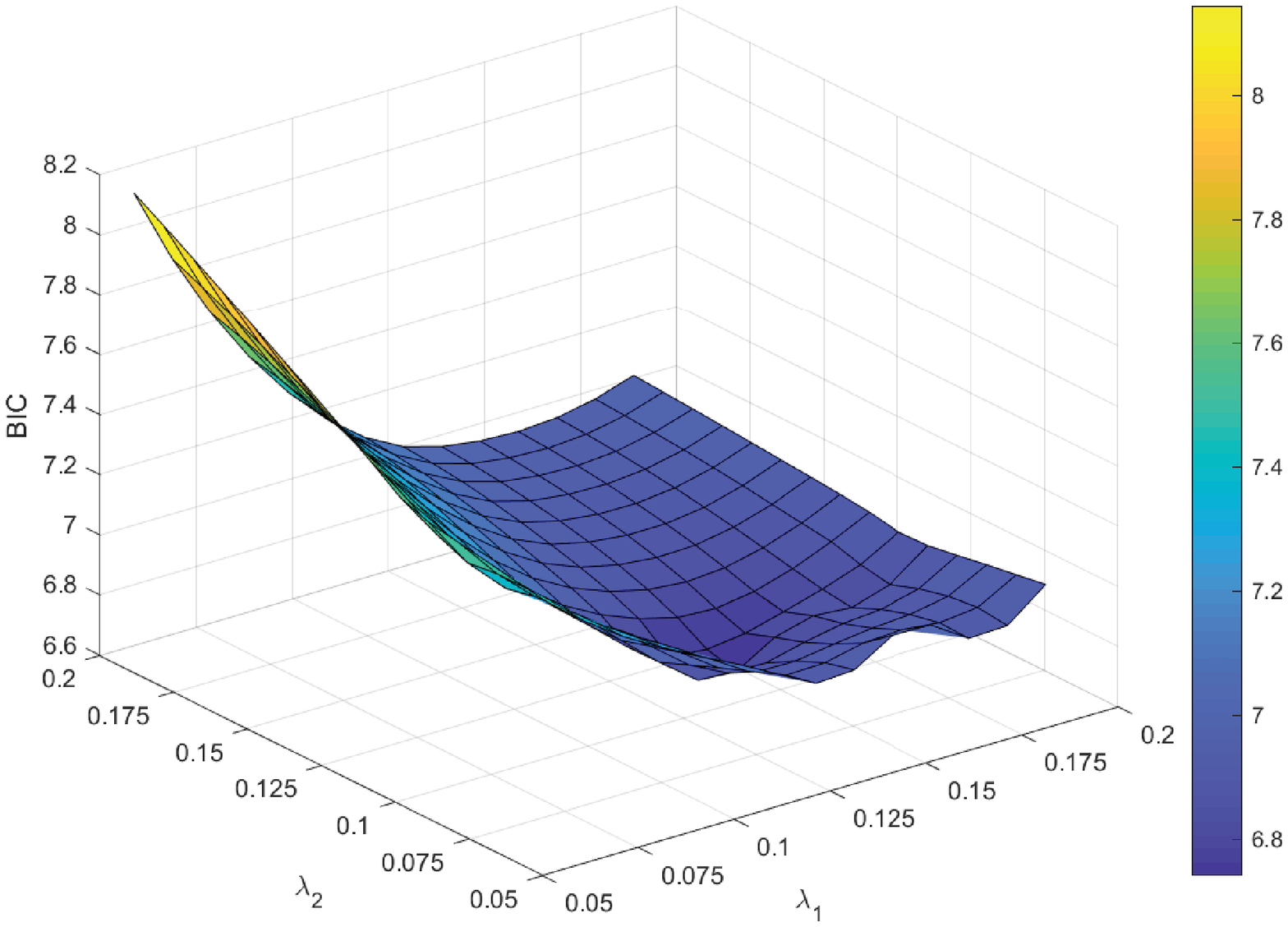}\\
  \caption{Simulation: BIC as a function of $\lambda_1$ and $\lambda_2$}\label{BIC_curve}
\end{figure}

\clearpage
\begin{figure}[H]
  \centering
  \includegraphics[width=1.0\textwidth, angle=0]{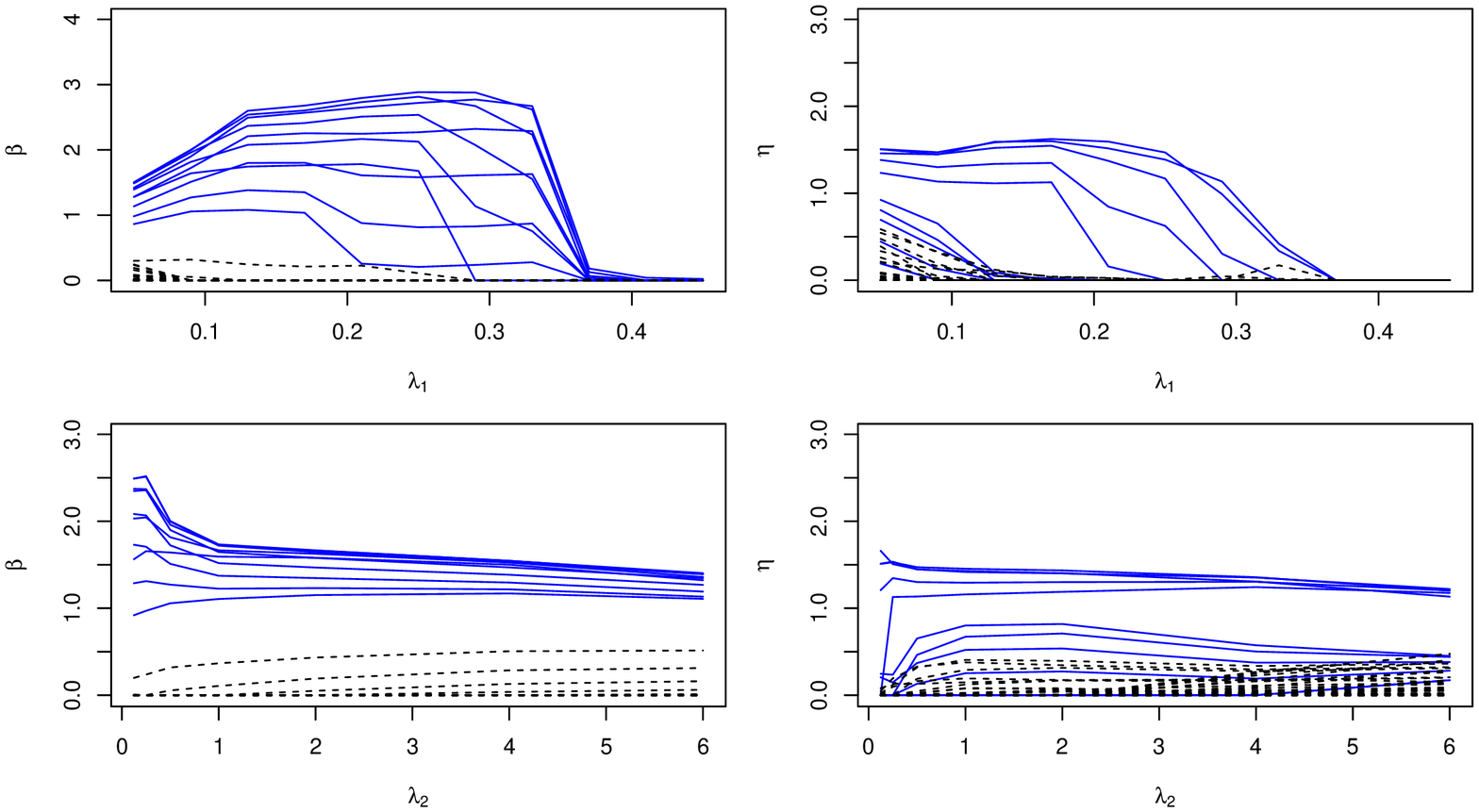}\\
  \caption{Simulation: parameter paths for one replicate under the linear model with MAF setting M1 and correlation structure AR(0.3). The blue solid lines represent the first ten true positives, and the black dashed lines represent the true negatives.}\label{para_path}
\end{figure}

\begin{center}
\begin{table}[h]
\caption{Simulation results under the AFT model with MAF setting M1. In each cell, mean (sd) based on 500 replicates.} \label{Scenario13-18}
\centering
\renewcommand{\tabcolsep}{0.2pc} % enlarge column spacing
\begin{tabular}{llllllll}
\hline
 & M:TP  &  M:FP  & I:TP  & I:FP  & RSSE  & RSE  & Cstat \\
\hline
\multicolumn{8}{c}{ AR(0.3) }\\
MA & 0.8(1.7) & 40.3(38.0) & 2.8(2.8) & 95.5(77.6) & 14.05(4.77) & 26.37(19.93) & 0.74(0.05) \\
HierMCP & 13.5(1.9) & 38.3(6.3) & 0.8(0.9) & 0.5(0.8) & 9.72(0.57) & 15.84(1.81) & 0.81(0.03) \\
SMCP & 4.8(4.8) & 6.5(13.7) & 9.1(6.6) & 31.9(15.7) & 8.54(0.56) & 3.24(0.80) & 0.85(0.03) \\
Proposed & 19.2(1.2) & 1.2(6.7) & 33.1(6.1) & 6.1(4.3) & 2.99(1.09) & 2.29(0.72) & 0.93(0.02) \\
\hline
\multicolumn{8}{c}{ AR(0.5) }\\
MA & 1.8(2.7) & 55.8(38.9) & 4.7(4.0) & 127.3(75.0) & 71.88(62.71) & 173.12(166.62) & 0.56(0.10) \\
HierMCP & 12.9(1.8) & 40.6(6.9) & 0.9(1.0) & 0.9(0.9) & 10.13(0.69) & 17.32(2.16) & 0.80(0.03) \\
SMCP & 6.5(4.9) & 7.9(12.9) & 8.9(7.2) & 34.4(16.3) & 8.36(0.72) & 3.50(0.91) & 0.85(0.03) \\
Proposed & 19.3(1.3) & 0.4(2.3) & 34.2(4.9) & 5.4(4.0) & 2.82(1.02) & 2.18(0.58) & 0.93(0.01) \\
\hline
\multicolumn{8}{c}{ Band1 }\\
MA & 1.1(1.9) & 41.6(42.5) & 2.5(2.6) & 92.7(80.5) & 13.71(4.68) & 25.49(20.59) & 0.72(0.08) \\
HierMCP & 13.5(1.7) & 37.1(5.7) & 0.7(0.7) & 0.4(0.8) & 9.66(0.51) & 15.60(1.67) & 0.81(0.03) \\
SMCP & 4.6(4.7) & 5.1(12.0) & 9.2(6.5) & 28.5(15.6) & 8.51(0.65) & 3.20(0.92) & 0.86(0.03) \\
Proposed & 19.3(1.2) & 0.4(1.4) & 33.4(5.4) & 6.1(4.9) & 2.92(0.90) & 2.25(0.76) & 0.93(0.01) \\
\hline
\multicolumn{8}{c}{ Band2 }\\
MA & 1.8(2.3) & 59.6(41.1) & 5.5(3.9) & 131.5(73.7) & 100.86(90.94) & 245.78(233.22) & 0.54(0.07) \\
HierMCP & 12.8(1.9) & 42.0(8.2) & 1.2(1.0) & 0.6(0.8) & 10.19(0.76) & 17.70(2.15) & 0.80(0.03) \\
SMCP & 9.4(5.2) & 17.8(18.5) & 9.3(6.5) & 35.8(14.1) & 8.10(0.83) & 4.00(1.01) & 0.85(0.03) \\
Proposed & 19.0(2.3) & 1.6(7.1) & 33.2(7.5) & 5.5(4.9) & 2.97(1.46) & 2.24(0.90) & 0.93(0.02) \\
\hline
\multicolumn{8}{c}{ LD(0.3) }\\
MA & 1.5(2.6) & 48.4(42.2) & 4.1(3.5) & 103.0(78.6) & 17.58(8.62) & 37.39(34.40) & 0.70(0.09) \\
HierMCP & 13.4(2.0) & 50.4(8.2) & 0.6(0.9) & 0.2(0.6) & 10.58(0.81) & 18.96(2.47) & 0.80(0.03) \\
SMCP & 4.4(4.5) & 4.7(11.7) & 11.7(8.1) & 21.3(11.2) & 8.33(0.88) & 2.98(0.67) & 0.86(0.03) \\
Proposed & 19.2(1.6) & 0.3(1.5) & 34.0(5.7) & 4.6(3.2) & 2.79(1.04) & 2.11(0.68) & 0.93(0.01) \\
\hline
\multicolumn{8}{c}{ LD(0.5) }\\
MA & 2.1(3.1) & 57.5(39.7) & 7.3(4.9) & 121.8(70.1) & 49.40(40.45) & 125.38(118.24) & 0.58(0.10) \\
HierMCP & 12.6(2.1) & 55.4(9.6) & 0.8(0.8) & 0.3(0.6) & 11.18(0.90) & 21.26(2.79) & 0.78(0.07) \\
SMCP & 6.8(5.7) & 8.0(13.8) & 13.8(8.6) & 22.7(12.9) & 7.87(1.02) & 3.16(1.04) & 0.85(0.08) \\
Proposed & 19.4(1.1) & 0.2(1.8) & 34.1(4.7) & 4.8(3.9) & 2.78(0.89) & 2.10(0.60) & 0.93(0.02) \\
\hline
\end{tabular}
\end{table}
\end{center}

\begin{center}
\begin{table}[h]
\caption{Simulation results under the AFT model with MAF setting M2. In each cell, mean (sd) based on 500 replicates.} \label{Scenario19-24}
\centering
\renewcommand{\tabcolsep}{0.2pc} % enlarge column spacing
\begin{tabular}{llllllll}
\hline
 & M:TP  &  M:FP  & I:TP  & I:FP  & RSSE  & RSE  & Cstat \\
\hline
\multicolumn{8}{c}{ AR(0.3) }\\
MA & 2.1(2.7) & 60.5(28.0) & 5.4(3.8) & 157.2(67.9) & 145.73(130.23) & 360.20(327.86) & 0.55(0.04) \\
HierMCP & 13.6(1.9) & 34.5(6.7) & 1.2(1.1) & 0.7(0.9) & 9.56(0.60) & 15.53(2.01) & 0.82(0.03) \\
SMCP & 5.8(4.8) & 9.4(15.1) & 4.1(2.7) & 81.9(20.0) & 8.68(0.37) & 3.47(0.57) & 0.82(0.05) \\
Proposed & 18.8(2.0) & 8.1(17.1) & 30.1(9.9) & 6.5(4.1) & 3.52(1.74) & 2.66(1.07) & 0.92(0.03)
\\
\hline
\multicolumn{8}{c}{ AR(0.5) }\\
MA & 3.2(3.5) & 68.6(21.8) & 7.8(4.9) & 172.3(47.2) & 184.51(107.73) & 451.07(267.52) & 0.53(0.04) \\
HierMCP & 12.8(1.8) & 37.9(7.4) & 1.2(1.2) & 0.9(0.9) & 10.05(0.75) & 17.06(2.41) & 0.80(0.03) \\
SMCP & 7.3(5.0) & 12.4(18.1) & 3.8(3.3) & 80.6(20.1) & 8.56(0.53) & 3.60(0.55) & 0.82(0.03) \\
Proposed & 18.9(2.2) & 6.5(19.5) & 31.9(9.4) & 5.6(4.9) & 3.22(1.80) & 2.37(0.97) & 0.92(0.03) \\
\hline
\multicolumn{8}{c}{ Band1 }\\
MA & 2.2(2.8) & 61.6(33.5) & 5.1(3.8) & 151.4(71.4) & 135.71(124.35) & 330.96(314.37) & 0.52(0.05) \\
HierMCP & 13.8(1.7) & 33.5(6.0) & 1.3(1.1) & 0.7(0.7) & 9.45(0.55) & 15.20(1.91) & 0.82(0.03) \\
SMCP & 6.3(4.7) & 8.0(13.0) & 4.0(3.2) & 79.8(19.9) & 8.62(0.42) & 3.40(0.51) & 0.84(0.02) \\
Proposed & 18.9(1.7) & 14.8(28.5) & 28.3(11.3) & 6.3(4.6) & 3.87(1.99) & 2.82(1.22) & 0.91(0.03) \\
\hline
\multicolumn{8}{c}{ Band2 }\\
MA & 3.8(3.6) & 64.9(24.8) & 8.8(4.7) & 167.1(51.5) & 207.33(179.97) & 519.71(455.48) & 0.53(0.04) \\
HierMCP & 12.6(1.8) & 38.8(8.1) & 1.6(1.3) & 1.0(1.0) & 10.11(0.75) & 17.74(2.17) & 0.80(0.03) \\
SMCP & 8.0(4.5) & 12.8(17.4) & 3.6(3.2) & 79.5(21.2) & 8.56(0.49) & 3.71(0.62) & 0.82(0.03) \\
Proposed & 18.5(2.8) & 11.3(24.1) & 29.4(11.5) & 5.5(4.0) & 3.67(2.08) & 2.65(1.26) & 0.91(0.03) \\
\hline
\multicolumn{8}{c}{ LD(0.3) }\\
MA & 1.5(2.6) & 48.4(42.2) & 4.1(3.5) & 103.0(78.6) & 17.58(8.62) & 37.39(34.40) & 0.70(0.09) \\
HierMCP & 13.4(2.0) & 50.4(8.2) & 0.6(0.9) & 0.2(0.6) & 10.58(0.81) & 18.96(2.47) & 0.80(0.03) \\
SMCP & 3.9(4.1) & 3.0(8.4) & 11.8(8.3) & 20.6(10.6) & 8.36(0.87) & 2.91(0.63) & 0.86(0.03) \\
Proposed & 19.2(1.6) & 0.3(1.5) & 34.0(5.7) & 4.6(3.2) & 2.79(1.04) & 2.11(0.68) & 0.93(0.01) \\
\hline
\multicolumn{8}{c}{ LD(0.5) }\\
MA & 2.1(3.1) & 57.5(39.7) & 7.3(4.9) & 121.8(70.1) & 49.40(40.45) & 125.38(118.24) & 0.58(0.10) \\
HierMCP & 12.6(2.1) & 55.4(9.6) & 0.8(0.8) & 0.3(0.6) & 11.18(0.90) & 21.26(2.79) & 0.78(0.07) \\
SMCP & 6.7(5.6) & 7.6(13.6) & 13.8(8.6) & 22.6(12.9) & 7.89(1.00) & 3.15(1.04) & 0.85(0.08) \\
Proposed & 19.4(1.1) & 0.2(1.8) & 34.1(4.7) & 4.8(3.9) & 2.78(0.89) & 2.10(0.60) & 0.93(0.02) \\
\hline
\end{tabular}
\end{table}
\end{center}

\clearpage
\begin{figure}
\centering
\subfigure{\scalebox{0.55}{\includegraphics{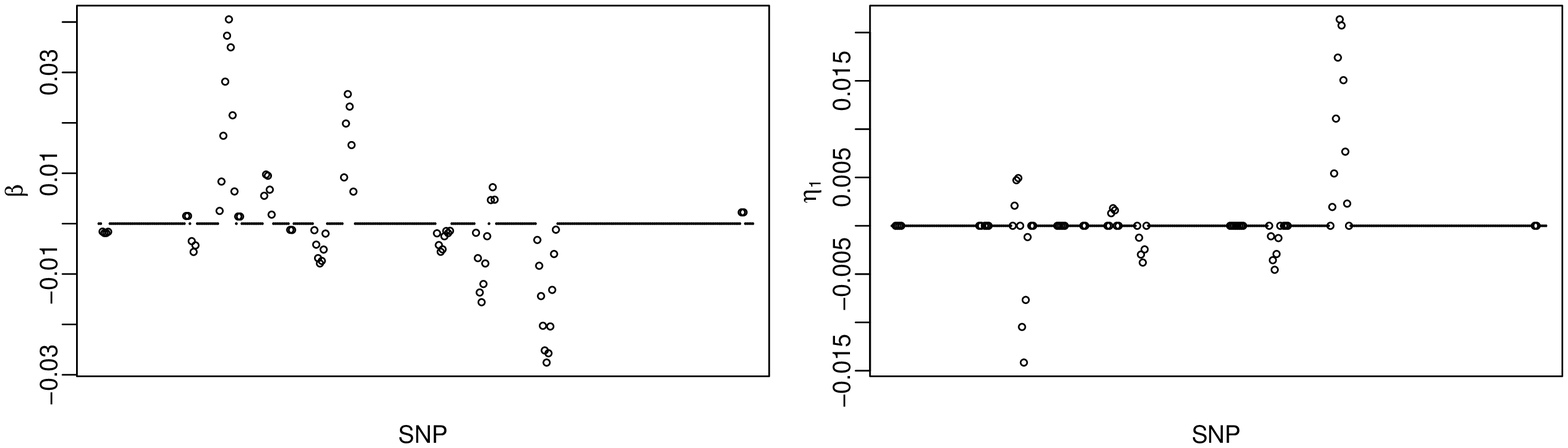}}
}
\subfigure{\scalebox{0.55}{\includegraphics{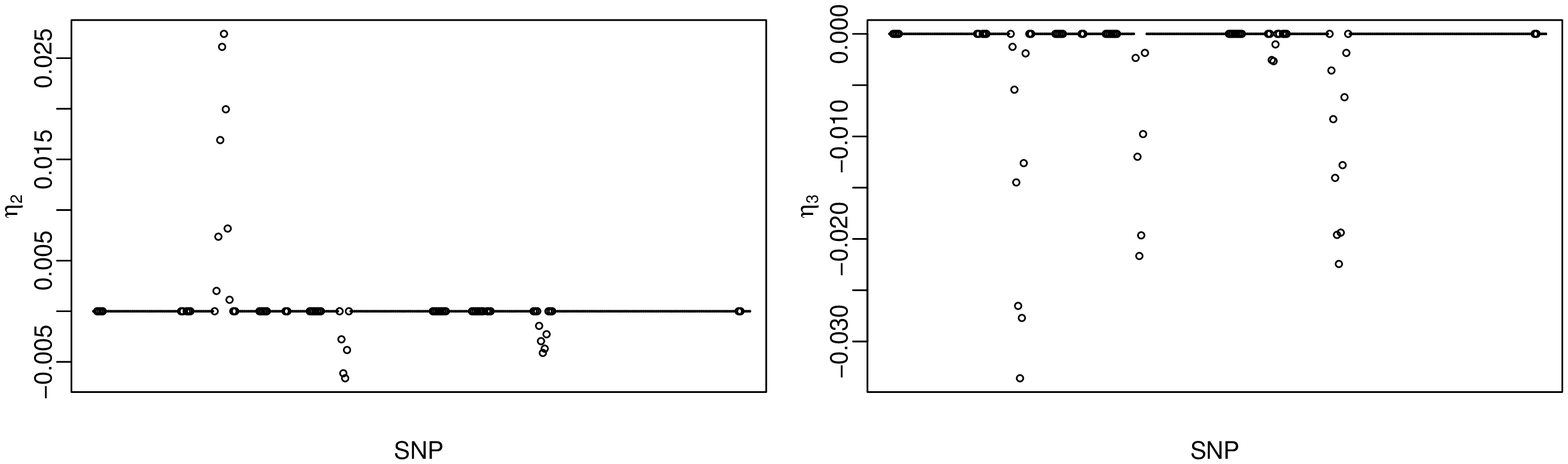}}
}
\subfigure{\scalebox{0.55}{\includegraphics{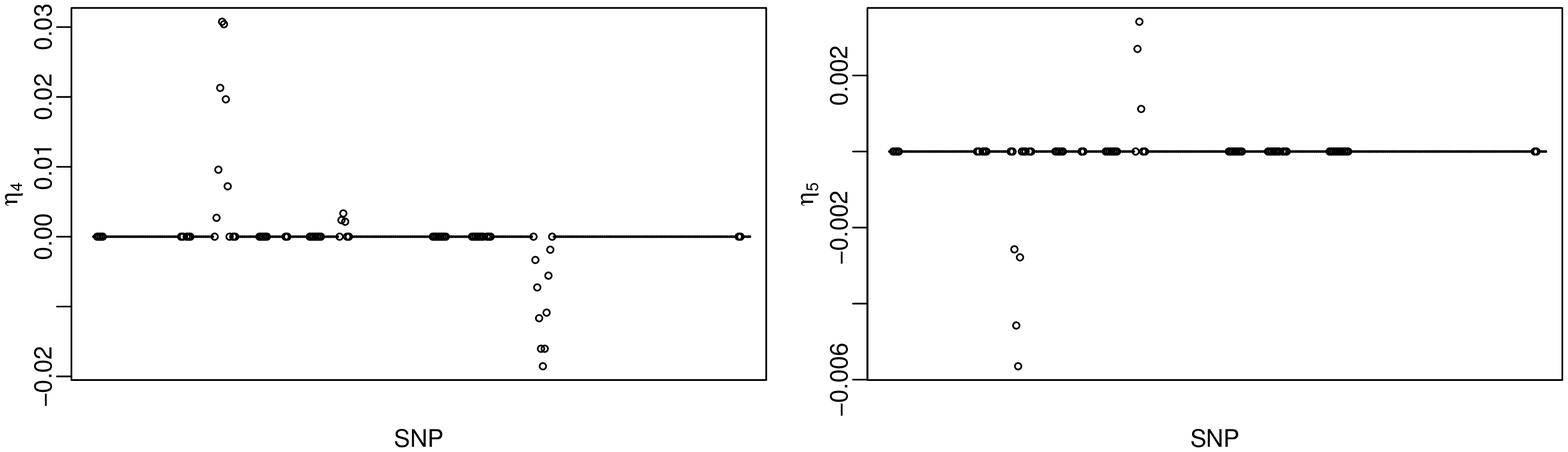}}
}
\subfigure{\scalebox{0.55}{\includegraphics{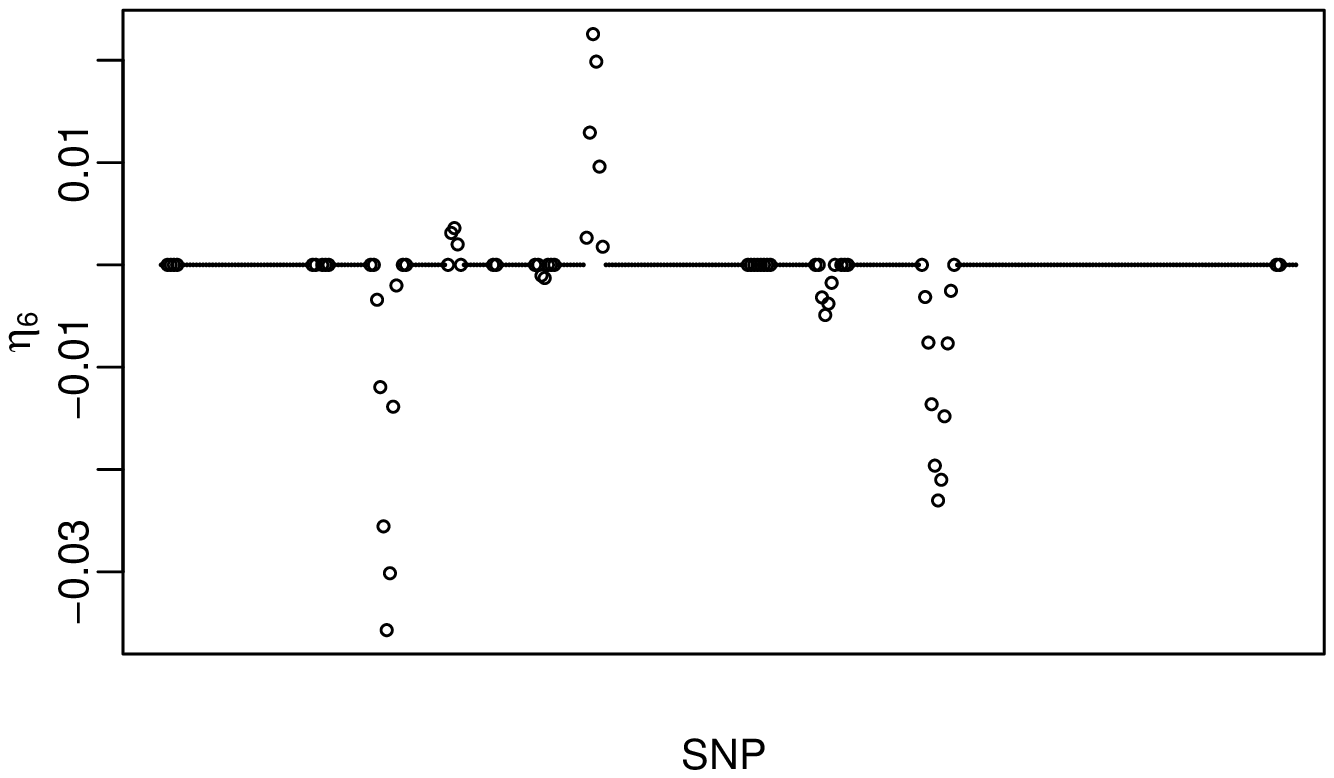}}
}\caption{Analysis of the GENEVA diabetes data (NHS/HPFS) using the proposed approach: identified main G effects and interactions.}
\label{Fig:GENEVA}
\end{figure}

\clearpage
\begin{figure}[H]
  \centering
  \includegraphics[width=1\textwidth, angle=0]{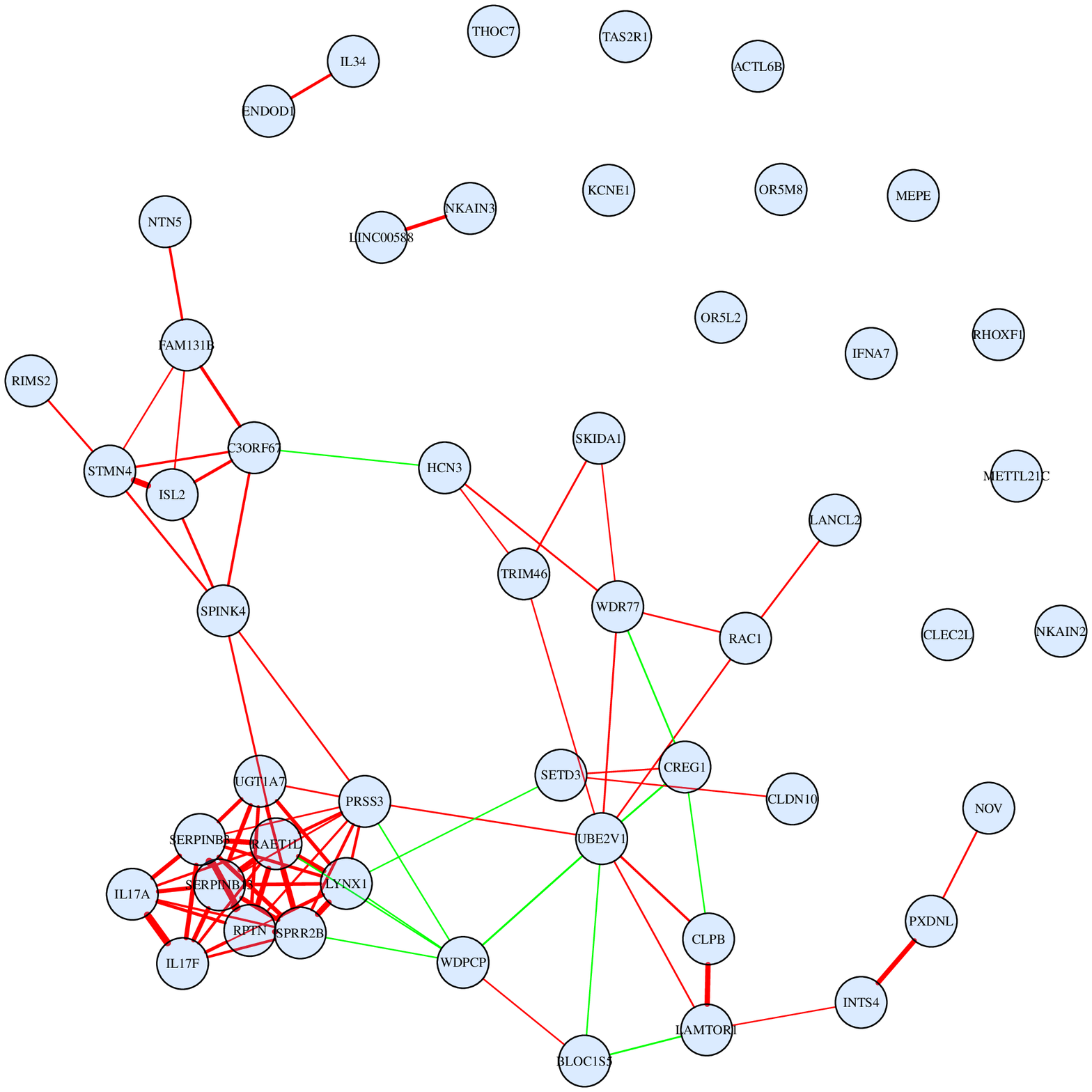}\\
  \caption{Analysis of the TCGA SKCM data using the proposed approach: identified main G effects. The edges between genes are defined based on the values of $a_{jl}$'s of the adjacency matrix $\bm{A}=(a_{jl})_{p \times p}$. Positive and negative connections are represented with red and green, respectively. The thickness (strength) of an edge is proportional to $|a_{jl}|$.}\label{Fig:SKCM}
\end{figure}

\clearpage
\begin{center}
\begin{table}[h]
\caption{Data analysis: numbers of main G effects and interactions identified by different approaches and their overlaps.} \label{Tab:overlap}
\renewcommand{\arraystretch}{1.4} % enlarge line spacing 0.3
\begin{tabular}{lccccccccccccccccccc}
\hline
GENEVA & \multicolumn{4}{c}{Main} & &\multicolumn{4}{c}{Interaction}  \\
\cline{2-5} \cline{7-10}
 & MA &HierMCP & SMCP & Proposed && MA &HierMCP & SMCP & Proposed \\
\hline
MA & 51 & 10 & 33 & 32 & & 57 & 0 & 31 & 0 \\
HierMCP & & 67 & 8 & 6 & & & 158 & 0 & 5 \\
SMCP & & & 41 & 30 & & & &  156 & 0\\
Proposed & & & &  71 & & & &   & 128 \\
 \hline
SKCM & \multicolumn{4}{c}{Main} & &\multicolumn{4}{c}{Interaction}  \\
\cline{2-5} \cline{7-10}
 & MA &HierMCP & SMCP & Proposed && MA &HierMCP & SMCP & Proposed \\
\hline
MA & 27 & 3 & 0 & 0 & & 21 & 0 & 0 & 0 \\
HierMCP & & 130 & 1 & 1 & & & 78 & 0 & 0 \\
SMCP & & & 39 & 15  & & & & 34 & 5\\
Proposed & & & & 50 & & & & & 44\\
 \hline
\end{tabular}
\end{table}
\end{center}

\end{document}